\documentclass[fleqn,usenatbib]{mnras}
\pdfoutput=1
\usepackage{newtxtext,newtxmath}

\usepackage[T1]{fontenc}
\usepackage{ae,aecompl}

\usepackage{graphicx}	
\usepackage{amsmath}	
\usepackage{amssymb}	
\usepackage{subfigure}

\title[Dynamics determine disc survival]{A tale of two clusters: dynamical history determines disc survival in Tr14 and Tr16 in the Carina Nebula}

\author[Reiter \& Parker]{
Megan Reiter,$^{1}$\thanks{E-mail: megan.reiter@stfc.ac.uk (MR)}
and Richard J. Parker$^{2}$\thanks{Royal Society Dorothy Hodgkin Fellow}
\\
$^{1}$UK Astronomy Technology Center, Royal Observatory Edinburgh, Blackford Hill, Edinburgh EH9 3HJ, UK\\
$^{2}$Department of Physics and Astronomy, The University of Sheffield,  Hicks Building, Hounsfield Road, Sheffield, S3 7RH, UK\\
}

\date{Accepted XXX. Received YYY; in original form ZZZ}

\pubyear{2019}

\begin{document}
\label{firstpage}
\pagerange{\pageref{firstpage}--\pageref{lastpage}}
\maketitle

\begin{abstract}
Understanding how the birthplace of stars affects planet-forming discs is important for a comprehensive theory of planet formation. Most stars are born in dense star-forming regions where the external influence of other stars, particularly the most massive stars, will affect the survival and enrichment of their planet-forming discs. Simulations suggest that stellar dynamics play a central role in regulating how external feedback affects discs, but comparing models to observations requires an estimate of the initial stellar density in star-forming regions. Structural analyses constrain the amount of dynamical evolution a star-forming region has experienced;  regions that maintain substructure and do not show mass segregation are likely dynamically young, and therefore close to their birth density. In this paper, we present a structural analysis of two clusters in the Carina Nebula, Tr14 and Tr16. We show that neither cluster shows evidence for mass segregation or a centrally concentrated morphology, suggesting that both regions are dynamically young. This allows us to compare to simulations from \citet{nicholson2019} who predict disc survival rates in star-forming regions of different initial densities. The surviving disc fractions in Tr14 and Tr16 are consistent with their predictions (both are $\sim 10$\%), supporting a growing body of evidence that the star-forming environment plays an important role in the survival and enrichment of protoplanetary discs. 
\end{abstract}

\begin{keywords}
stars: kinematics and dynamics -- stars: pre-main-sequence -- protoplanetary discs -- open clusters and associations: general
\end{keywords}



\section{Introduction}

  Most stars do not form in isolation; instead they form in aggregates of a few to many stars where feedback from nearby cluster members may alter their formation and evolution.
  Efforts to dissect the role of the star formation environment tend to focus on nearby regions ($d<1$~kpc) where individual sources can most readily be resolved and studied.
  These regions are predominantly forming low-mass stars. 
  However, fossil evidence in Solar System meteorites suggests that at least one dying high-mass star enriched the proto-Solar nebula/disc, providing the short-lived radioisotopes that play an important role in the geochemical evolution of terrestrial planets \citep[e.g.,][]{cameron1977,grimm1993,hester2004}. 
  Moreover, observations suggest a cluster mass function $dN/dM \sim M^{-2}$ \citep{lada2003,chandar2009,fall2010} which implies that $>1/2$ of all stars form in clusters more massive than the Orion Nebula Cluster (ONC), the prototypical high-mass star-forming region.

  Different theories for the formation of high-mass stars predict distinct cluster architectures.
  Under competitive accretion, high-mass stars form in the deepest part of the gravitational potential well, aided by high gas densities that enhance accretion rates \citep[e.g.,][]{bonnell2001}. However, feedback from these same massive stars may erase any observable difference in the spatial distribution \citep{parker_dale_2017}. For example, we wouldn't necessarily expect competitive accretion to give primordial mass segregation \citep{bonnell_davies1998}.
  Turbulent core models \citep[e.g.,][]{mckee2003} describe a formation pathway more analogous to that developed for isolated low-mass stars; high-mass stars (and their host (sub)clusters) form from high-density clumps in substructured, hierarchical clouds \citep{kruijssen2012}. 
Both formation scenarios predict a high degree of \emph{initial} spatial substructure, but this can rapidly evolve into a more centrally-concentrated morphology via rapid dynamical evolution in regions with high stellar densities, smoothing initially clumpy distributions and fostering mass segregation \citep[e.g.,][]{allison2009,allison2010,yu2011,parker2014}.
  Alternatively, a star-forming region could form in a smooth, very dense configuration and rapidly evolve into a less dense association due to residual gas expulsion \citep[e.g.,][]{tutukov1978,lada1984,goodwin1997,goodwin2006,baumgardt2007}, although it is currently unclear how this would result in the spatial and kinematic substructure observed in stellar associations \citep{wright2014,wright2016,ward2018}. 
 Dynamical processing does not create or enhance substructure, so its persistence may be taken as evidence for dynamical youth \citep{scally_clarke2002,goodwin_whitworth2004,parker2012,parker2014}.

Motivated by the observed composition of our own Solar System, \citet{adams2001} estimated the size of the Sun's birth cluster, arguing for an aggregate with $N \approx 2000 \pm 1100$ stars. Most star-forming regions in the immediate vicinity of the Sun are less populous, with only $N \approx 300 - 1000$ members. In these small star-forming regions, tidal effects and external irradiation from other members may not play a significant role in shaping the nascent planetary systems \citep[e.g.,][]{adams2004,adams2006}. 
The key parameter is the median local stellar density -- a high local density can lead to the creation of free-floating planets and orbital disruption even if the total number of stars is low \citep{parker2012a}. 
Such small star-forming regions may not include stars massive enough to produce short-lived radioisotopes within the lifetime of discs and planet formation \citep[though see][who argue that the only limit to the mass of a star that can form is the mass of the cloud itself]{elmegreen2006,parker2007}, meaning that enrichment may only occur in more massive star-forming regions \citep[][show that if low-mass star-forming regions can form massive stars, then radioistopic enrichment is possible]{nicholson2017}.  

If we impose a limit on the mass of the most massive star that can form in a given region \citep{weidner2006}, then a star-forming region with a few thousand stars will only produce one or two stars $>$20\,M$_\odot$.  By way of example, a $\sim 25$~M$_{\odot}$ star lives for $\sim 7.5$~Myr, longer than the dissipation time for the majority of protoplanetary discs around low-mass stars \citep{haisch2001,richert2018}. 
Higher-mass stars, typically found in more populous star-forming regions, evolve faster and may produce radioisotopes more efficiently in the later stages of their pre-supernova evolution, providing earlier enrichment \citep{knoedlseder1996,voss2012}. 

While mass-loss from evolved high-mass stars will pollute the cluster environment, it is unclear what fraction of protostellar discs will be enriched. 
Prior to their explosive deaths, high-mass stars will also bathe the cluster with energetic radiation that will rapidly photoevaporate the gas component of planet forming-discs around low-mass stars \citep[e.g.,][]{storzer1999,scally_clarke2001,adams2004,fatuzzo2008,mann2010,mann2014,haworth2018,winter2018,nicholson2019}. 
Nevertheless, planets are prevalent \citep[e.g.,][]{dressing2013}, with terrestrial planets mostly likely to be found around low-mass stars \citep[e.g.,][]{howard2012,mulders2015}.
Those enriched with short-lived radioisotopes from dying high-mass stars may have conditions more favorable for habitability \citep[see, e.g.,][]{lugaro2018}. 
Understanding the role of the cluster environment is thus an essential part of a comprehensive theory of planet formation.

Recent work from \citet{lichtenberg2016,nicholson2019} suggests that stellar dynamics in the natal cluster play a central role in the survival and enrichment of planet-forming discs. 
  In high density regions, rapid dynamical evolution brings low-mass stars close to the high-mass stars where their discs are quickly destroyed.
  Lower density environments evolve more slowly, allowing low-mass stars to spend more time at a safe distance from the destructive radiation of the high-mass stars.

  The Gaia revolution is underway \citep[e.g.,][]{gaia2018}, providing parallaxes and proper motions of billions of stars and reinvigorating dynamical studies of young open clusters \citep[e.g.,][]{damiani2017_VelOB2,franciosini2018,roccatagliata2018}. 
  At the distance of the typical high-mass star-forming region ($\gtrsim 2$~kpc), Gaia measurements are most reliable for the brightest, and therefore highest mass cluster members \citep[e.g.,][]{kuhn2019}. 
  Most radial velocity surveys target modestly sized clusters in the Solar neighborhood \citep[e.g.,][]{furesz2008,tobin2009,foster2015,cottaar2015,dario2017}, with only a few studies targeting high-mass regions \citep[e.g.,][]{damiani2017,karnath2019}.
  As a result, few constraints exist for truly high-mass regions \citep[e.g.,][]{wright2014,wright2016}.

In the absence of comprehensive kinematic studies of high-mass star-forming regions, statistical diagnostics provide insight into the formation pathway and dynamical state of young clusters. 
A variety of approaches to estimate structure, morphology, and clustering have been proposed \citep[e.g.,][]{cartwright2004,allison2009,maschberger2011,kuhn2014,buckner2019}. 
These metrics provide a framework for comparing an ensemble of clusters, useful to constrain their probable formation and evolution pathways \citep[e.g.,][]{kuhn2014,kuhn2015}. 
For individual clusters, structure diagnostics, especially when combined with density estimates, provide strong constraints on the dynamical history of the cluster \citep{parker2012,parker2014,parker2014c}. 
This is of particular interest for high-mass clusters as simulations suggest that their dynamics determine the integrated feedback affecting planet-forming discs around nearby low-mass stars, and thus dictate their survival and enrichment.

In this paper, we consider the central clusters of the Carina Nebula, Tr14 and Tr16.
These clusters sample the two archetypal morphologies produced by theories of high-mass star-formation -- centrally-concentrated and hierarchical, respectively. 
Both clusters are likely at the same distance \citep{turner1980,walborn1995,smith2006_energy,hur2012}, and close enough to feasibly observe both low- and high-mass stars \citep{smith2006_distance,hur2012}. 
Tr14 appears to be centrally concentrated \citep{ascenso2007,sana2010,kuhn2014} and a few authors report tentative evidence for mass-segregation (e.g., \citealt{sana2010}; \citealt{buckner2019} although see \citealt{ascenso2007}). 
In contrast, Tr16 is hierarchical, with considerable substructure and no clear cluster center \citep[e.g.,][]{wolk2011,kuhn2014}. 
The total stellar content and average densities of the two clusters are similar \citep[see, e.g., Table~7 in][]{wolk2011}. 
However, stellar densities in the centrally-concentrated core of Tr14 are an order of magnitude higher than in Tr16 \citep[][]{ascenso2007,sana2010} which has no clear cluster center. 
Multiple age indicators suggest that Tr16 is slightly older than Tr14 \citep[$\sim 3$~Myr, compared with $\sim 1$~Myr for Tr14; see e.g.,][]{walborn1995,preibisch2011_hawki,getman2014}.
Together, Tr14 and Tr16 sample the key cluster morphologies to test the role of cluster dynamics in the disc survival prior to the onset of the supernova era.

We quantify substructure and mass-segregation in these two clusters in order to estimate their dynamical histories. 
This allows us to estimate the impact of external feedback on the planet-forming discs around nearby low-mass stars by comparing numerical simulations from  \citet{nicholson2019} with the surviving disc fraction in the two clusters from \citet{preibisch2011}. 
In doing so, we constrain the role of cluster density in determining the destiny of protoplanetary discs within them.

\section{Point Source Catalogs}

We combine point-source catalogs in the literature to perform the structural analysis of Carina. 
Our primary focus is the two central clusters, Tr14 and Tr16; however, we include observations of the larger star-forming complex
in order to compare the structure of the entire region to that determined in the
clusters. 
We compile multiple surveys to maximize spatial coverage of Carina and to sample a broader range of stellar masses.
We use the K-band magnitude as a proxy for mass, selecting the near-IR filter to minimize the effects of uneven extinction in Tr14 and Tr16. 
In tests of synthetic clusters with and without extinction, 
\citet{parker2012_rho_oph} found that mass-segregation diagnostics recover strong signals of mass segregation, but the statistical significance is somewhat reduced in the presence of extinction.

For a census of O- and B-type stars in Carina, we use the list of known Carina members and new spectroscopic confirmations from \citet{alexander2016}.
To provide sources with a broad range of masses in and around Tr14 and Tr16, we use the photometric study of \citet{hur2012}. 
Those authors identify cluster members using a combination of
proper motions, 
spectral types, 
reddening characteristics, 
X-ray emission, and 
near-IR excess.
For both of these catalogs, we cross-match with 2MASS \citep{skrutskie2006} to obtain K~mags for each source.
We include stars in Tr14 detected with AO-assisted observations by 
\citet{sana2010} who identify cluster members in the high-density core of Tr14 based on their position in near-IR color space. 
We also use two IR catalogs of young stellar objects (YSOs) that cover the larger Carina star-forming region produced as part of the Chandra Carina Complex Project \citep{townsley2011}.
First, we use the near-IR point source catalog from 
\citet{preibisch2011} who used associated X-ray emission to distinguish young cluster members from background contaminants with similar IR colors. 
Second, we include candidate intermediate-mass YSOs identified by \citet{povich2011} based on their IR SEDs. 
In total, this provides a catalog of 9236 point sources distributed over $\sim 1.5^{\circ} \times 2.5^{\circ}$ (see Figure~\ref{fig:full_carina}).

Each of these surveys cover a different footprint, providing uneven sensitivity and spatial coverage. 
To ensure that this does not alter the results of the structural analysis, we repeat the analysis using only the X-ray-selected sample of low-mass stars from \citet{kuhn2014}. 
We cross-match the \citet{kuhn2014} catalog with the near-IR data from \citet{preibisch2011} to provide K~mags. 
We compare the distribution of low-mass stars with the O- and B-type stars from \citet{alexander2016}. 
This provides a slightly smaller sample of $\sim 1300$ objects in each cluster.

\section{Structure diagnostics}\label{s:results}

In this Section we apply three different diagnostics for quantifying the spatial structure of star-forming regions to the full Carina region, as well as the Tr14 and Tr16 clusters individually. We first briefly describe each of the diagnostics before applying them to the observational data.

\subsection{Description of diagnostics}

\subsubsection{The $\mathcal{Q}$-parameter}\label{ss:qparam}

The $\mathcal{Q}$-parameter \citep{cartwright2004,cartwright2009,lomax2011,jaffa2017} quantifies whether a star-forming region has a substructured or smooth morphology by comparing the mean edge length of a minimum spanning tree connecting all of the points, $\bar{m}$ with the mean edge length of a complete graph of the distribution, $\bar{s}$:
\begin{equation}
    \mathcal{Q} = \frac{\bar{m}}{\bar{s}}.
\end{equation}
$\bar{m}$ is normalised by the following factor, which depends on the number of stars in the distribution, $N$, and the area $A$:
\begin{equation}
    \frac{\sqrt{NA}}{N - 1}.
\end{equation}
The area, $A$, is the area of a circle with radius $R$ centred on the region in question, and where $R$ is the radius of the region's outermost star from the centre. $\bar{s}$ is then normalised to the radius $R$ of this circle (and so $\mathcal{Q}$ is a dimensionless ratio). \citet{parker2018} shows that this is the most robust normalisation technique when determining $\mathcal{Q}$. 

In two dimensions, $\mathcal{Q} < 0.8$ indicates a substructured morphology, whereas $\mathcal{Q} > 0.8$ indicates a smooth, centrally concentrated distribution.

\subsubsection{The $\Lambda_{\rm MSR}$ mass segregation ratio}\label{ss:Lmsr}

The mass segregation ratio, $\Lambda_{\rm MSR}$ \citep{allison2009} provides a quantitative measure of mass segregation by comparing the minimum spanning tree length of randomly chosen subsets of stars in a star-forming region with the length of a minimum spanning tree of a chosen subset. In this case, we are interested in the high-mass stars, but $\Lambda_{\rm MSR}$ can be adapted to any subset of interest.

$\Lambda_{\rm MSR}$ is defined as
\begin{equation} 
\Lambda_{\rm MSR} = \frac{\langle l_{\rm average} \rangle} {l_{\rm massive}}
\end{equation}
where 
$\langle l_{\rm average} \rangle$ is the average edge length of the minimum spanning tree of many subsets of randomly chosen stars
and 
$l_{\rm massive}$ is the typical distance between high-mass stars. 

The $\Lambda_{\rm MSR}$ ratio is determined for subsets of the $N_{\rm MST}$ most massive (or brightest -- see below) stars. The initial choice for the number of stars in the subset is $N_{\rm MST} = 4$, $\Lambda_{\rm MSR}$ is calculated for this number of stars, increasing in increments of six stars until the number of stars in the subset is equal to the total number of stars in the dataset (and for this number of stars $\Lambda_{\rm MSR} = 1$ by definition). If a region is significantly mass-segregated  then $\Lambda_{\rm MSR} >> 1$, and numerical experiments show that values above 2 are generally not produced by random chance \citep{parker_goodwin_2015}. 

The determination of $\Lambda_{\rm MSR}$ requires no assumptions about the center of a star-forming region, or its morphology, and is a single metric that may be used to measure mass segregation in the full region, as well as Tr14 and Tr16. The technique can be applied to any scalar quantity; usually this is stellar mass, but for our dataset the K-band magnitudes are more reliable and we will use those in our analysis as a proxy for mass.

\subsubsection{Relative local surface density of the most massive stars}\label{ss:sigma}

\citet{maschberger2011} introduced another method to quantify the relative spatial distribution of massive stars by comparing the local surface density, $\Sigma$, around each star as a function of stellar mass. 

Following \citet{casertano1985} we compute the local surface density $\Sigma$, by determining the distance $d_n$ to the $n^{\rm th}$ nearest neighbor, such that 
\begin{equation}
    \Sigma = \frac{n-1}{\pi d_n^2}.
\end{equation} 
The choice of $n$ is somewhat arbitrary. Care must be taken to avoid low values of $n$ which would introduce a bias due to enhancements in local density from  binary or high-order multiple systems; similarly a high value of $n$ would wash out the effects of density enhancements inherent in a substructured star-forming region. We adopt $n = 10$ throughout this work \citep[see also][]{parker2014}.

Following \citet{parker2014}, we compute the local surface density ratio 
\begin{equation}
    \Sigma_{\mathrm{LDR}} = \frac{\tilde{\Sigma}_{10}}{\tilde{\Sigma}_{\rm all}},
\end{equation}
which compares
the local surface density of the 10 most massive stars in the star-forming region, 
$\tilde{\Sigma}_{10}$,
with the local surface density of all stars in the cluster, $\tilde{\Sigma}_{\rm all}$.  This method was designed to compare surface density and mass, but we will use K-band magnitudes instead of mass.

To gauge the significance of any difference between the median surface density of the subset of interest and the median surface density of the entire region, we perform a Kolmogorov-Smirnov (KS) test between the two populations and reject the null hypothesis that they share the same underlying parent distribution if the KS test returns a p-value $<0.01$.

 $\Sigma_{\rm LDR}$ measures whether a high-mass star resides in a higher-than-average density location within a star-forming region, and is distinct from the $\Lambda_{\rm MSR}$ mass segregation ratio, which measures the relative positions of the high-mass stars compared to low-mass stars. It is possible for a star-forming region to display a high  $\Sigma_{\rm LDR}$ ratio, but a low  $\Lambda_{\rm MSR}$ ratio, and vice versa \citep[see][]{parker_goodwin_2015}.

\subsection{The full Carina region}
\begin{figure*}
  \centering
\hspace*{-1.5cm}\subfigure[]{\label{fig:full_Carina-a}\rotatebox{0}{\includegraphics[trim=1cm 0cm 5cm 1cm,scale=0.4]{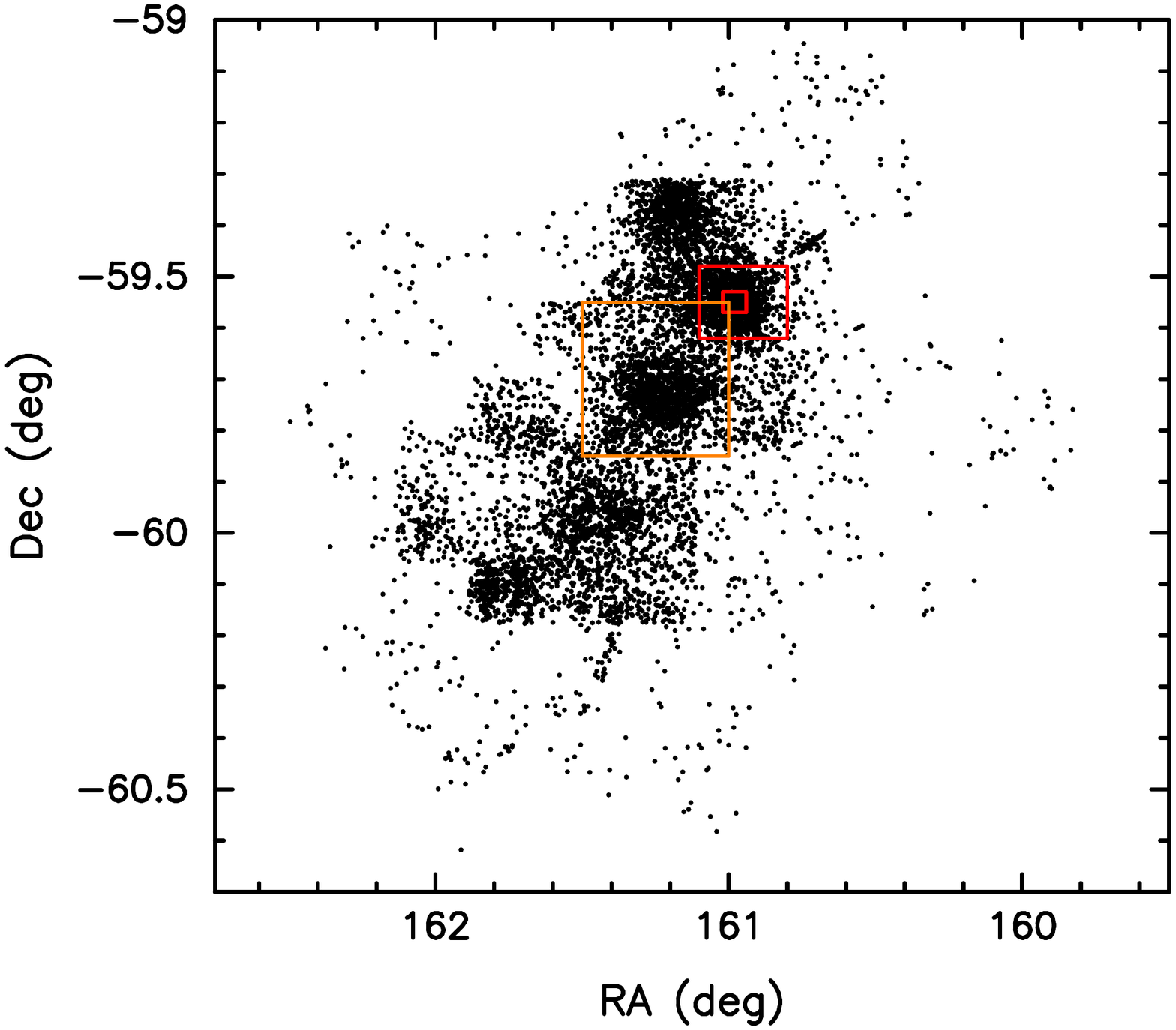}}}
\subfigure[]{\label{f1}\rotatebox{0}{\includegraphics[trim=1cm 0cm 5cm 1cm,scale=0.4]{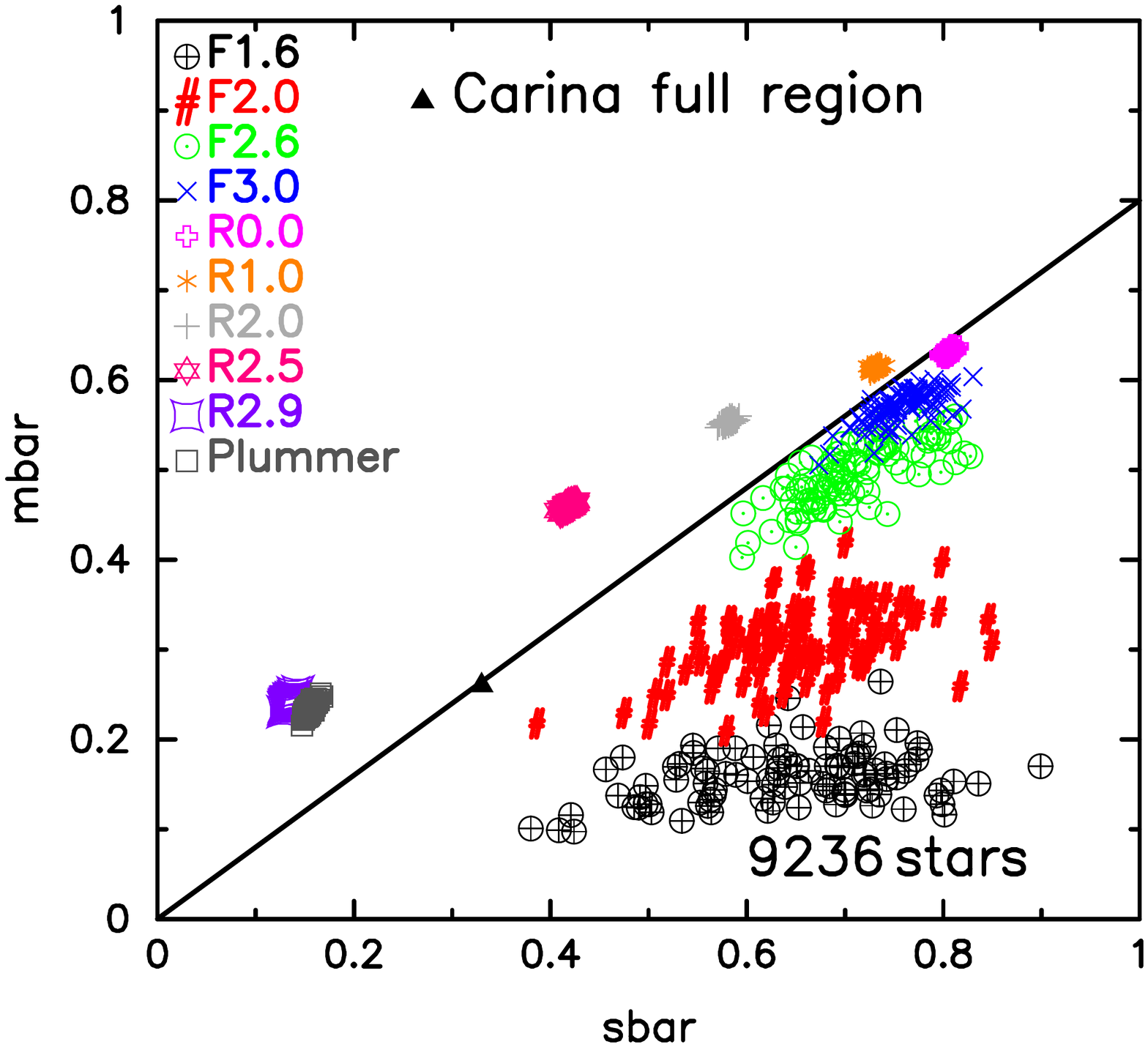}}} 
\hspace*{-0.85cm}\subfigure[]{\label{f2}\rotatebox{0}{\includegraphics[trim=2cm 0cm 5cm 1cm,scale=0.37]{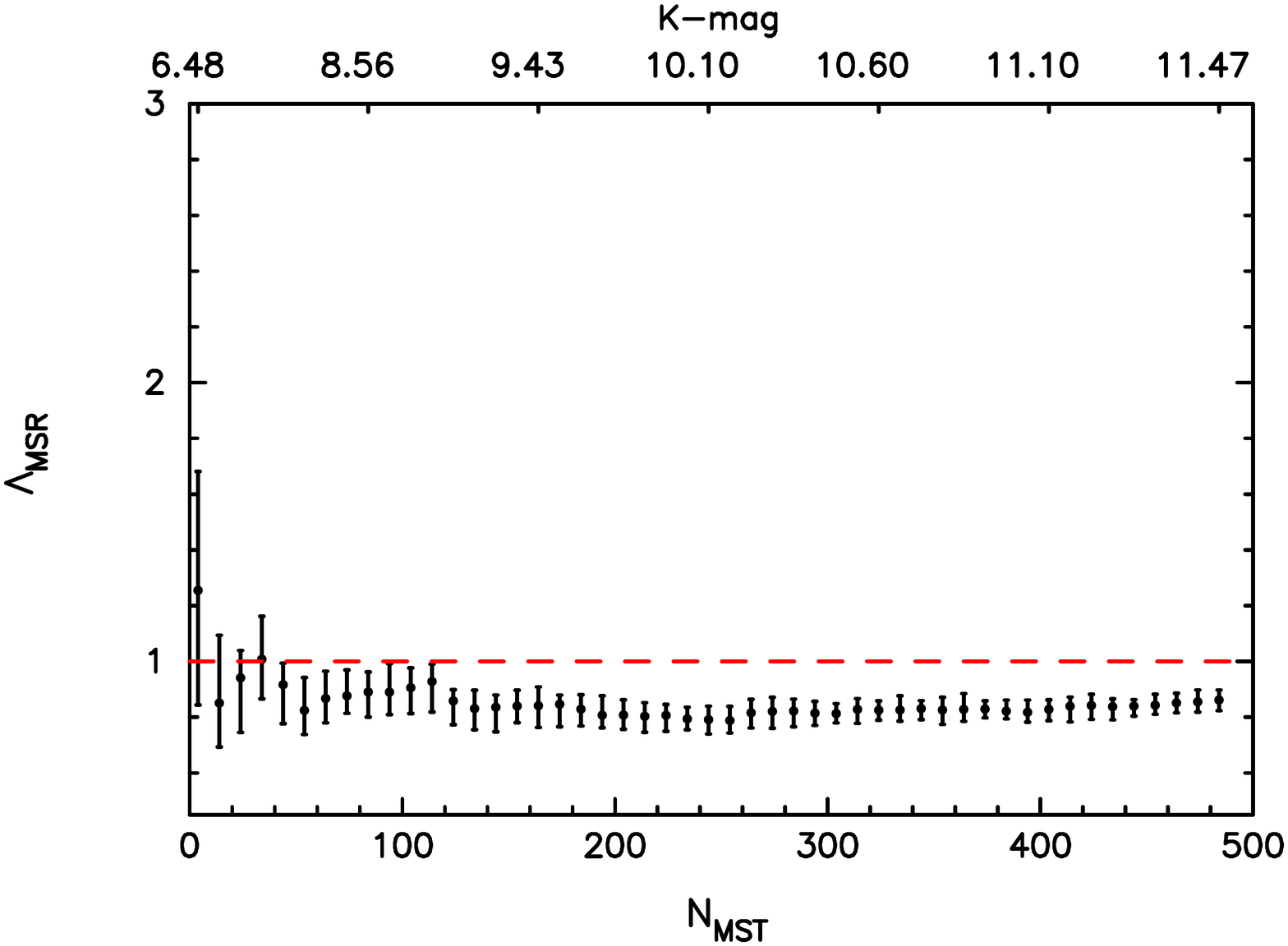}}}
\hspace*{0.5cm}\subfigure[]{\label{f3}\rotatebox{0}{\includegraphics[trim=1cm 0cm 5cm 1cm,scale=0.4]{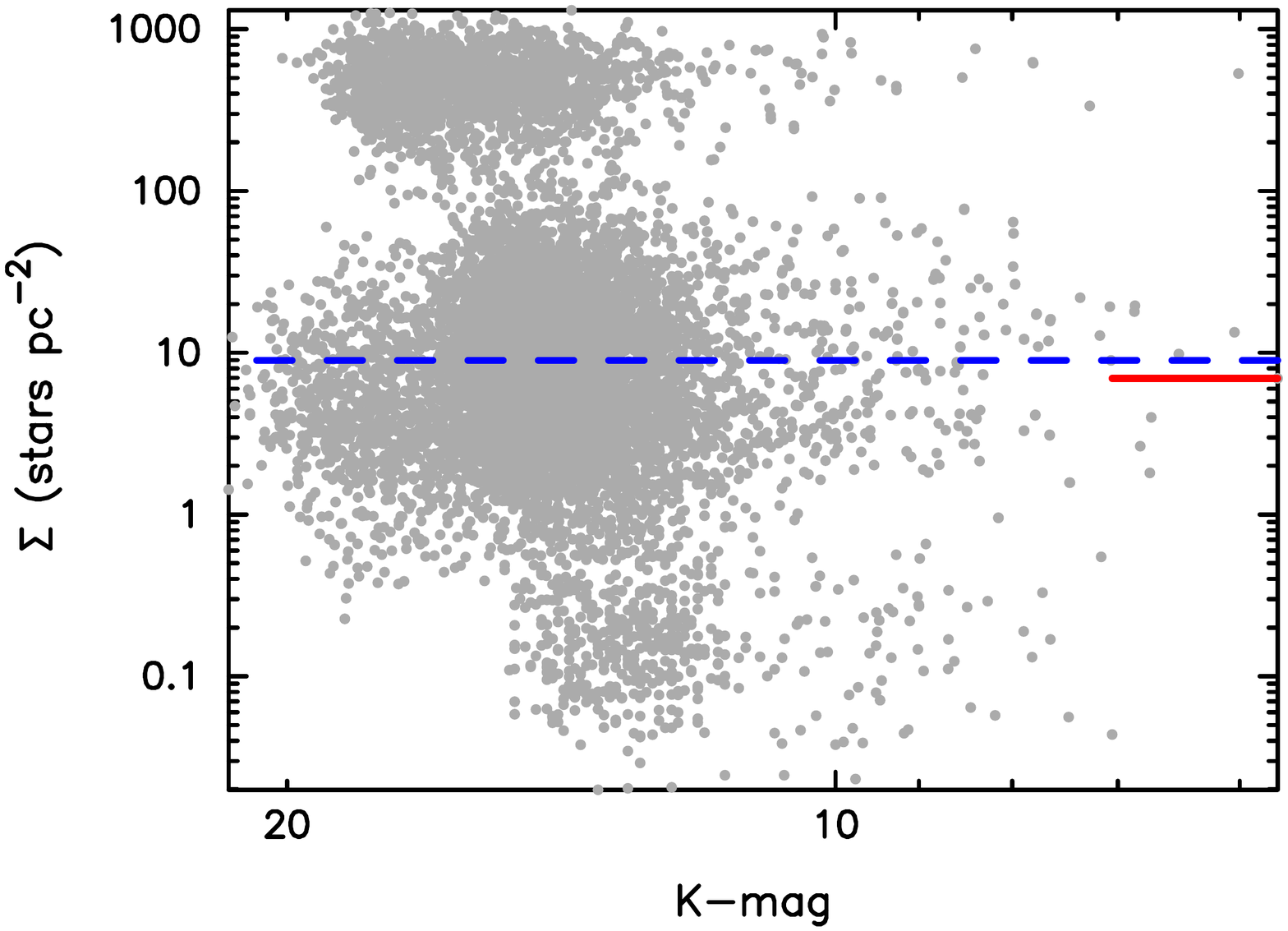}}}
\caption{Structural analysis of the entire dataset. Panel (a) shows the sub-regions in the following figures. The largest red outline box is Trumpler 14; the small red box within this is a zoom-in of the densest part of Trumpler 14; the orange outline box is Trumpler 16. In panel (b) we show the \citet{cartwright2009} plot, which plots the $\bar{m}$ and $\bar{s}$ components used to calculate the $\mathcal{Q}$-parameter against each other. The datum for the entire Carina region is shown by the solid black triangle, and for reference we show 100 realisations each of synthetic star-forming regions with various fractal dimensions, $D$, (from a high degree of substructure [$D = 1.6$] to smooth [$D=3.0$] -- indicated as F1.6 - F3.0 in panel (b)), or centrally concentrated regions with different density profiles (uniform [$n \propto r^0$] to very centrally concentrated [$n \propto r^{-2.9}$] -- indicated as R0.0 - R2.9 in panel (b), as well as a Plummer profile). In panel (c) we show the $\Lambda_{\rm MSR}$ ratio as a function of the $N_{\rm MST}$ brightest stars. The K-band magnitude of the least bright object enclosed in a sample of $N_{\rm MST}$ stars is indicated on the top axis. $\Lambda_{\rm MSR} = 1$ (no mass segregation) is shown by the dashed red line. In panel (d) we show the surface density $\Sigma$ for each star as a function of its K-band magnitude. The median surface density for the Carina region is shown by the blue dashed line, and the median surface density for the ten brightest (OB) stars is shown by the solid red line.
}\label{fig:full_carina} 
\end{figure*}

If we consider the full Carina region (Fig.~\ref{fig:full_carina}a), for the spatial distribution we measure $\bar{m} = 0.26$ and $\bar{s} = 0.33$, giving $\mathcal{Q} = 0.79$. If we place this datum on the \citet{cartwright2009} $\bar{m} - \bar{s}$ plot (Fig.~\ref{fig:full_carina}b), we see that the morphology of the region is not consistent with a simple geometry. In this figure we show 100 realisations each of fractal distributions where we increase the fractal dimension from $D =1.6$ (highly substructured) to $D = 3.0$ (smooth). We also show 100 realisations each of centrally-concentrated clusters with morphologies described by a density profile of the form $n \propto r^{-\alpha}$, where we increase the degree of central concentration from uniform ($\alpha = 0$) to significantly concentrated ($\alpha = 2.9$). We also show clusters with a \citet{plummer1911} profile. 

The neutral value for $\mathcal{Q}$ for the entire Carina complex can be explained if the star-forming region is transitioning from a fractal distribution to a centrally-concentrated cluster, but in this case is more likely to be due to the superposition of different structures within the same field of view \citep{parker2012,parker_dale2015}.  

A plot of the mass segregation ratio, $\Lambda_{\rm MSR}$, against the $N_{\rm MST}$ subset of the brightest stars (Fig.~\ref{fig:full_carina}c) displays no deviation from $\Lambda_{\rm MSR} = 1$, suggesting that the region is not mass segregated. 

Similarly, the stars with the brightest K-band magnitudes in the full sample are not found in regions of higher than average surface density (Fig.~\ref{fig:full_carina}d). 
 This plot readily shows the three different density regimes in the dataset\footnote{The origin of different density regimes in star-forming regions is the subject of much debate. \citet{larson1995} suggests they trace different scales of star formation, from core fragmentation on sub pc scales to clustering on pc scales to diffuse star formation on larger scales. Alternatively, it may simply indicate the dynamical evolution of multiscale star-formation \citep{kraus2008,kruijssen2012}.}; a diffuse component with density $\tilde{\Sigma} \sim  0.1$\,stars\,pc$^{-2}$, an average component for the star-forming region which has $\tilde{\Sigma} \sim 10 - 100$\,stars\,pc$^{-2}$ and a dense component (due to the different subclusters) which has  $\tilde{\Sigma} \sim 100 - 1000$\,stars\,pc$^{-2}$, consistent with other estimates \citep[see Table~7 in][]{wolk2011}. The median surface density for the entire region is $\tilde{\Sigma}_{\rm all} \sim 9$\,stars\,pc$^{-2}$, and the ten brightest stars have a median density of  $\tilde{\Sigma}_{\rm 10} = 7$\,stars\,pc$^{-2}$.  A KS-test between the two populations returns a p-value of 0.67 that they share the same underlying parent distribution. We therefore conclude that the most massive stars in Carina are not found in locations with different stellar surface densities to the average stars.

\subsection{Tr14}

We now focus on the subscluster Tr14, and analyse two different datasets. The first is our compilation of sources, and the second is the dataset from \citet{kuhn2014}, which was also analysed by \citet{buckner2019}. 

\subsubsection{Compilation data}

\begin{figure*}
  \begin{center}
\hspace*{-1.5cm}\subfigure[]{\label{fig:Tr14_comp-a}\rotatebox{0}{\includegraphics[trim=1cm 0cm 5cm 1cm,scale=0.4]{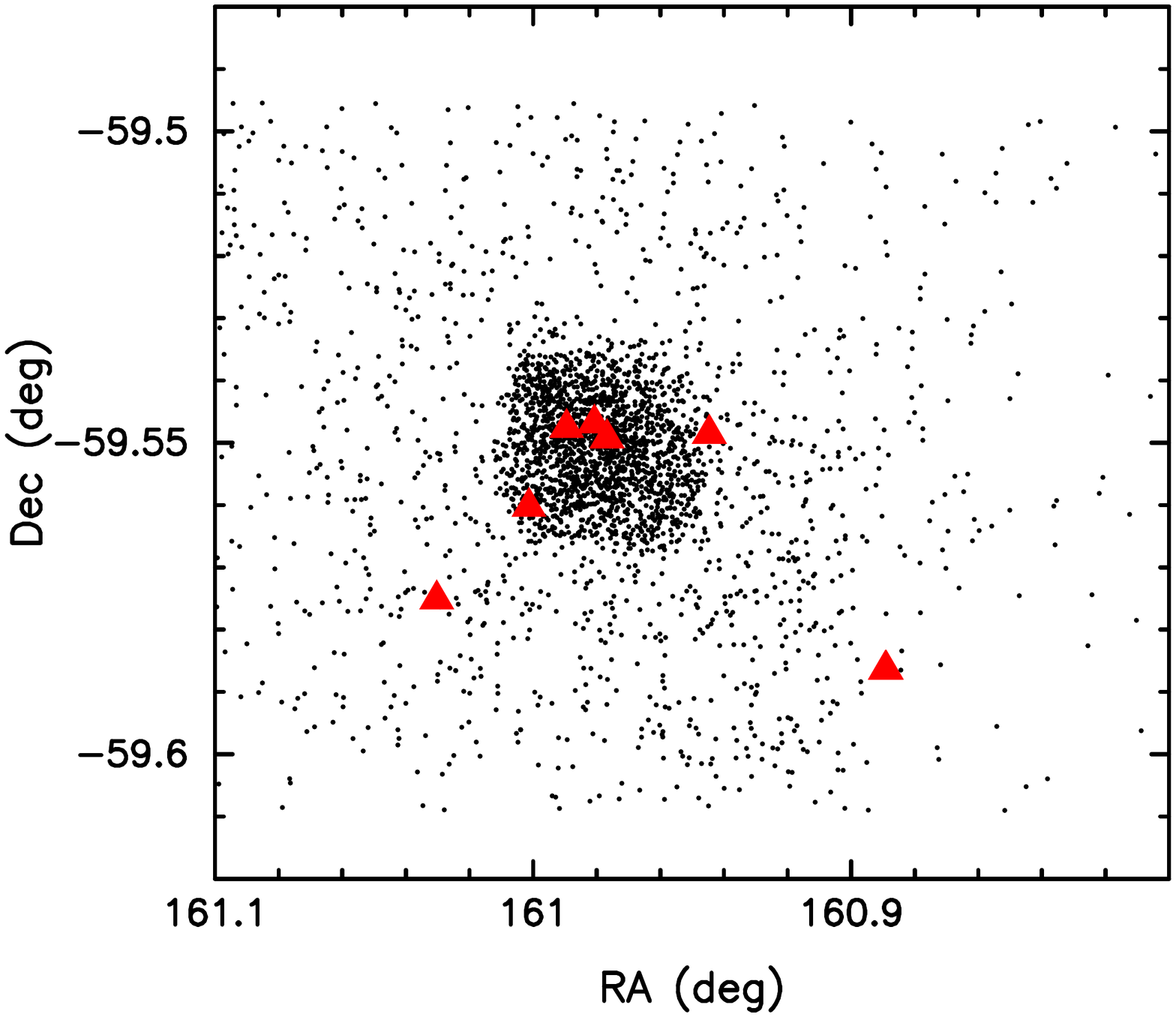}}}
\subfigure[]{\label{fig:Tr14_comp-b}\rotatebox{0}{\includegraphics[trim=1cm 0cm 5cm 1cm,scale=0.4]{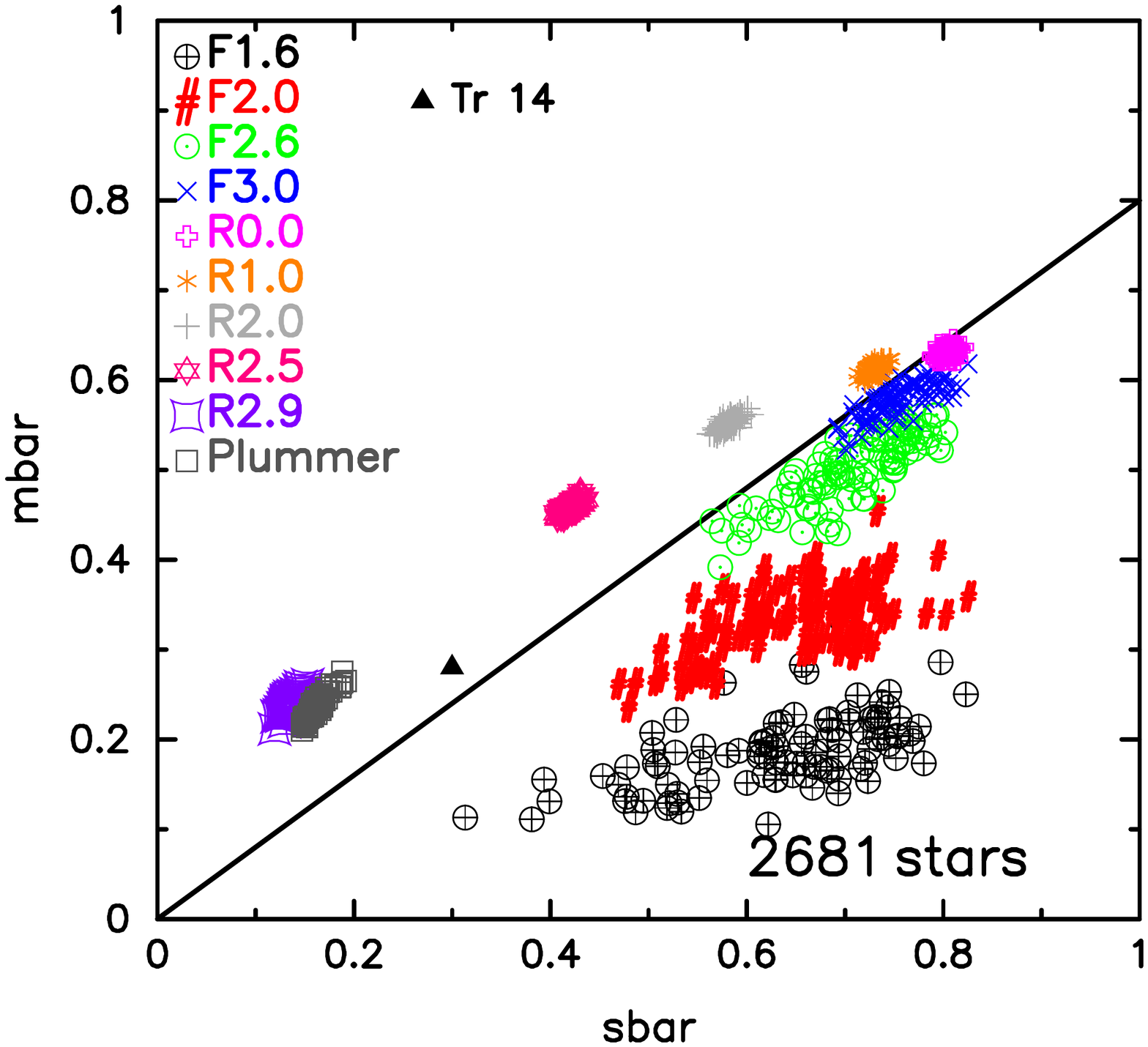}}} 
\hspace*{-0.85cm}\subfigure[]{\label{fig:Tr14_comp-c}\rotatebox{0}{\includegraphics[trim=2cm 0cm 5cm 1cm,scale=0.37]{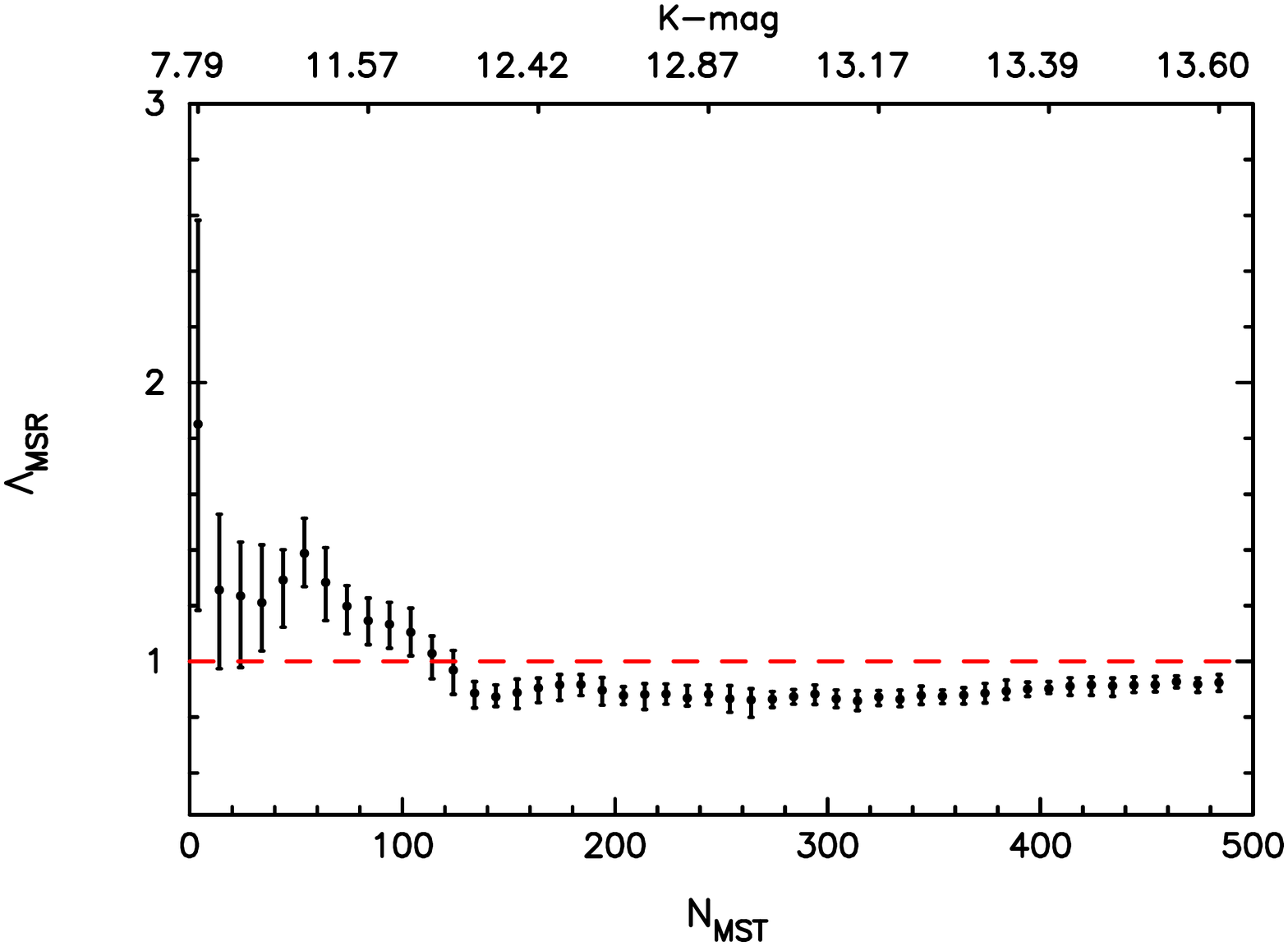}}}
\hspace*{0.5cm}\subfigure[]{\label{fig:Tr14_comp-d}\rotatebox{0}{\includegraphics[trim=1cm 0cm 5cm 1cm,scale=0.4]{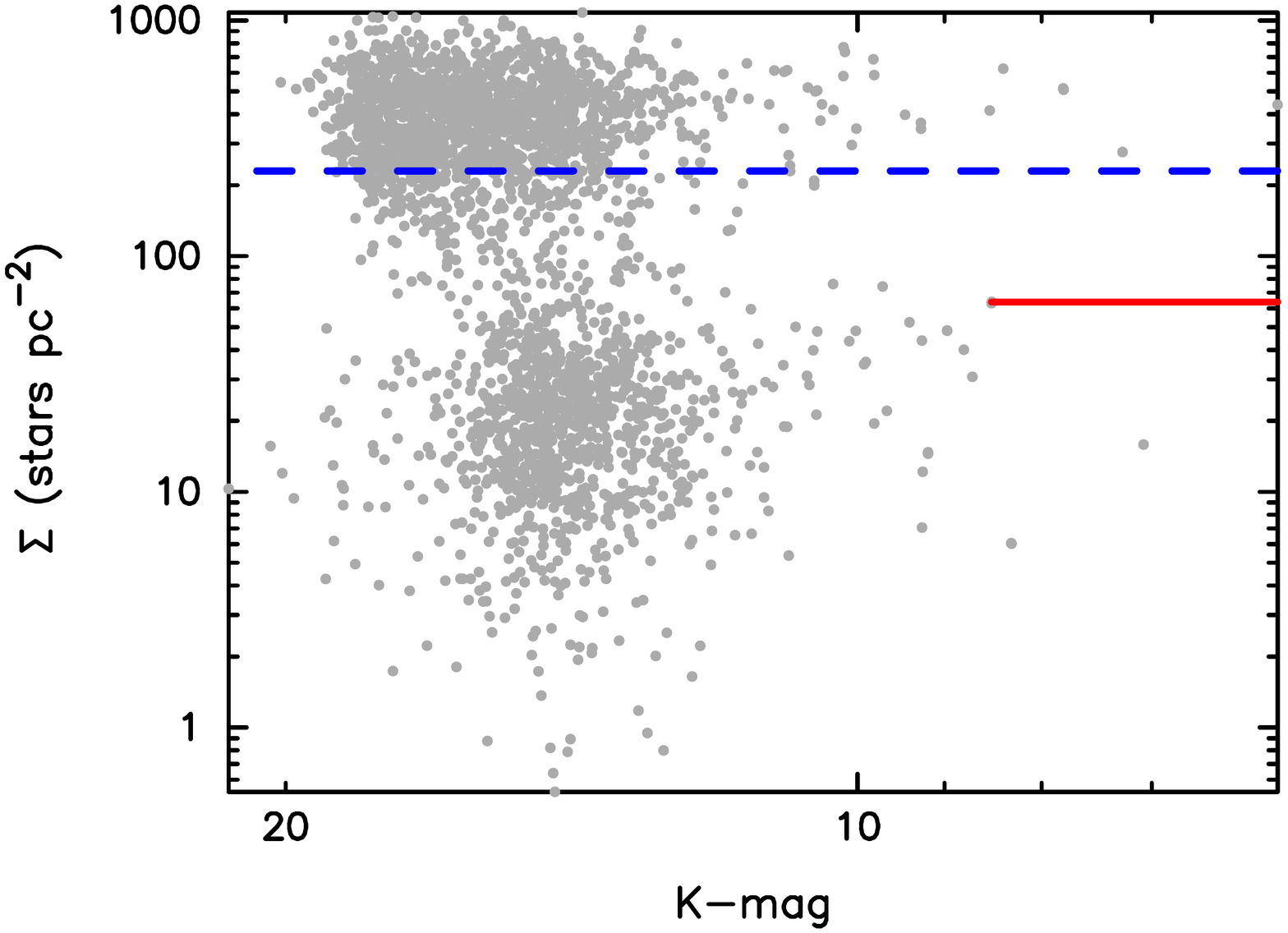}}}
\caption[bf]{Structural analysis of the Trumpler 14 dataset. In panel (a) the brightest stars are shown by the red points.  In panel (b) we show the \citet{cartwright2009} plot, which plots the $\bar{m}$ and $\bar{s}$ components used to calculate the $\mathcal{Q}$-parameter against each other. The datum for Tr14 is shown by the solid black triangle, and for reference we show 100 realisations each of synthetic star-forming regions with various fractal dimensions, $D$, (from a high degree of substructure [$D = 1.6$] to smooth [$D=3.0$] -- indicated as F1.6 - F3.0 in panel (b)), or centrally concentrated regions with different density profiles (uniform [$n \propto r^0$] to very centrally concentrated [$n \propto r^{-2.9}$] -- indicated as R0.0 - R2.9 in panel (b), as well as a Plummer profile). In panel (c) we show the $\Lambda_{\rm MSR}$ ratio as a function of the $N_{\rm MST}$ brightest stars.  The K-band magnitude of the least bright object enclosed in a sample of $N_{\rm MST}$ stars is indicated on the top axis. $\Lambda_{\rm MSR} = 1$ (no mass segregation) is shown by the dashed red line. In panel (d) we show the surface density $\Sigma$ for each star as a function of its K-band magnitude. The median surface density for the Carina region is shown by the blue dashed line, and the median surface density for the ten brightest (OB) stars is shown by the solid red line.
}
\label{fig:Tr14_comp}
  \end{center}
\end{figure*}

In Fig.~\ref{fig:Tr14_comp} we show the results for the Tr14 subcluster, focusing on the area defined by the larger red outline box in Fig.~\ref{fig:full_Carina-a}. A zoomed-in view of this area is shown in Fig.~\ref{fig:Tr14_comp-a}. The ten brightest stars are shown by the red triangles; note that several of these systems are massive binaries and the points are superimposed. We also note that the contrast between the central region and the surrounding outskirts reflects differing spatial coverage and depths of the surveys used to make the combined catalog; however, we have re-analysed the central region, and used the independent dataset from \citet{kuhn2014} and our results are very similar (see Section~\ref{ss:kuhn_comp}).

We calculate a $\mathcal{Q}$-parameter of 0.94 for Tr14, where $\bar{m} = 0.28$ and $\bar{s} = 0.30$. Taken in isolation, $\mathcal{Q} = 0.94$ would suggest a smooth, centrally concentrated morphology. However, in Fig.~\ref{fig:Tr14_comp-b} we show the location of this datum on the \citet{cartwright2009} $\bar{m} - \bar{s}$ plot, which shows that the observational data is not consistent with a simple centrally-concentrated morphology. It may reflect the superposition of two different distributions \citep{parker_dale2015}, or it could represent a mid-point in the dynamical evolution of a substructured spatial distribution to a smoother one \citep[as dynamics always erase substructure,][]{scally_clarke2002,goodwin_whitworth2004,parker2014}. 

The $\Lambda_{\rm MSR}$ mass segregation ratio as a function of the $N_{\rm MST}$ stars in the chosen subset is shown in Fig.~\ref{fig:Tr14_comp-c}. Whilst several of the datapoints lie slightly above  $\Lambda_{\rm MSR} = 1$, none of them fulfill the additional criteria that $\Lambda_{\rm MSR} \geq 2$, which was suggested by \citet{parker_goodwin_2015} to alleviate "significant" deviations from unity that can be caused by random chance. We therefore posit that the full region is not mass segregated. 

The surface density of every star in the sample is plotted against K-band magnitude in Fig.~\ref{fig:Tr14_comp-d}. Due to the bimodal nature of the data coverage for Tr14, the central region appears to have a higher density ($\Sigma \sim 500$\,stars\,pc$^{-2}$) and the outer areas have a much lower density ($\Sigma \sim$ tens\,stars\,pc$^{-2}$).

The median density for the full Tr14 sample is $\tilde{\Sigma}_{\rm all} = 229$\,stars\,pc$^{-2}$, whereas the ten most massive stars have a lower density ($\tilde{\Sigma}_{\rm 10} = 67$\,stars\,pc$^{-2}$). However, a KS test between the two distributions returns $D = 0.16$ and a p-value = 0.9 that they share the same underlying parent distribution. The reason for this is that the massive stars are distributed over a wide range of stellar surface densities (for example, the star above the median sits in an area of local surface density of 276\,stars\,pc$^{-2}$).

In summary, the massive, or brightest stars in our Tr14 sample are not spatially distributed differently to the average stars in this (sub)cluster. Using a new clustering algorithm, \citet{buckner2019} find that the brightest stars in Tr14 are more clustered that lower-mass stars, using the dataset in \citet{kuhn2014}, whereas we find no evidence of preferential clustering of the most massive stars. We test whether this is due to our adoption of different samples in the following subsection.

\subsubsection{\citet{kuhn2014} data}\label{ss:kuhn_comp}

In order to test whether our results are dependent on the uneven sensitivity and spatial coverage of the catalogs we combine to sample the stellar distribution, we apply $\mathcal{Q}$, $\Lambda_{\rm MSR}$ and $\Sigma_{\rm LDR}$ using only the catalog of low-mass point sources from \citet{kuhn2014} and high-mass stars from \citet{alexander2016}. 

\begin{figure*}
  \begin{center}
\setlength{\subfigcapskip}{10pt}
\hspace*{-1.0cm}\subfigure[]{\label{fig:Tr14_kuhn-a}\rotatebox{0}{\includegraphics[trim=1cm 0cm 5cm 1cm,scale=0.25]{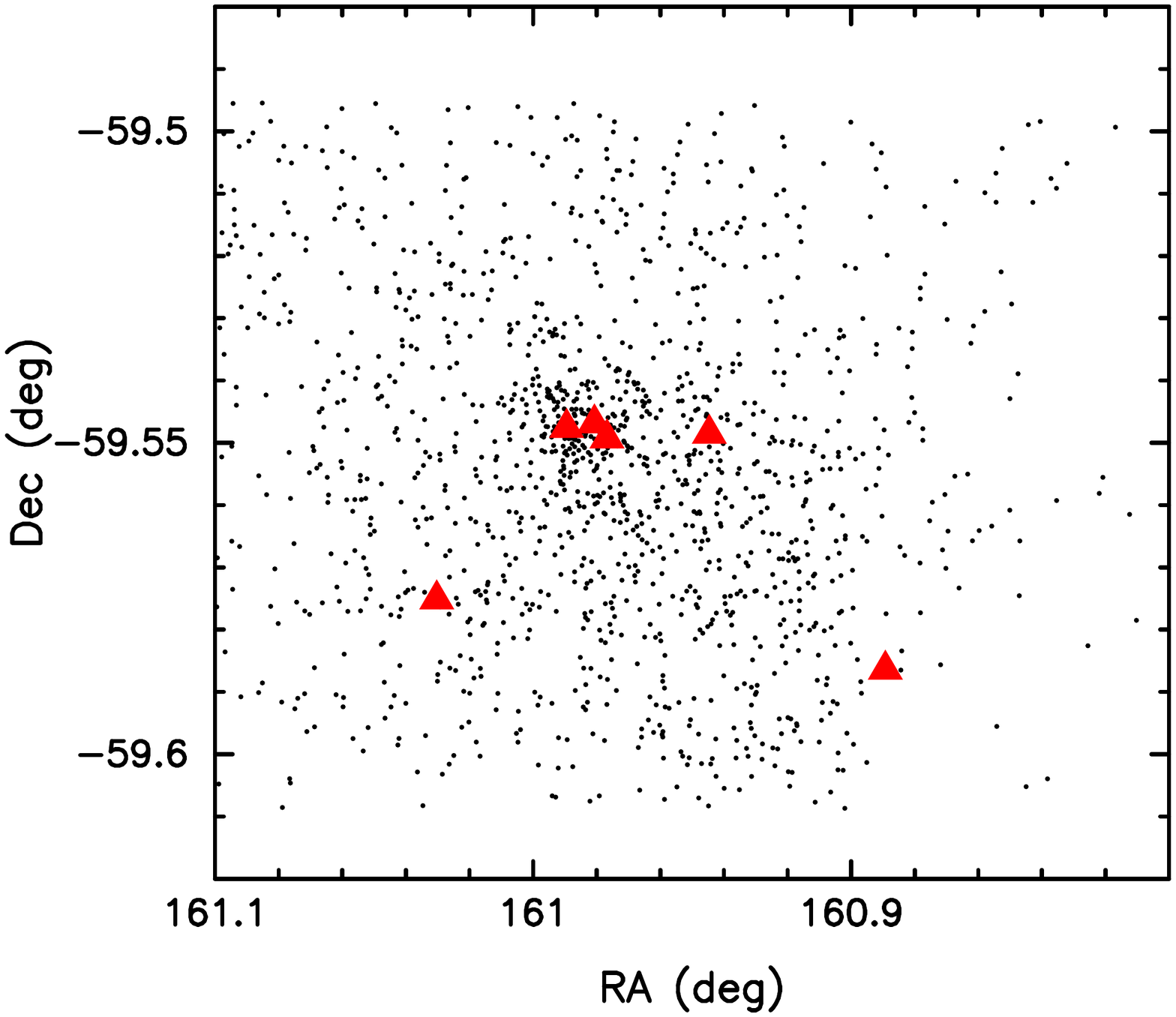}}}
\subfigure[]{\label{fig:Tr14_kuhn-b}\rotatebox{0}{\includegraphics[trim=2cm 0cm 5cm 1cm,scale=0.25]{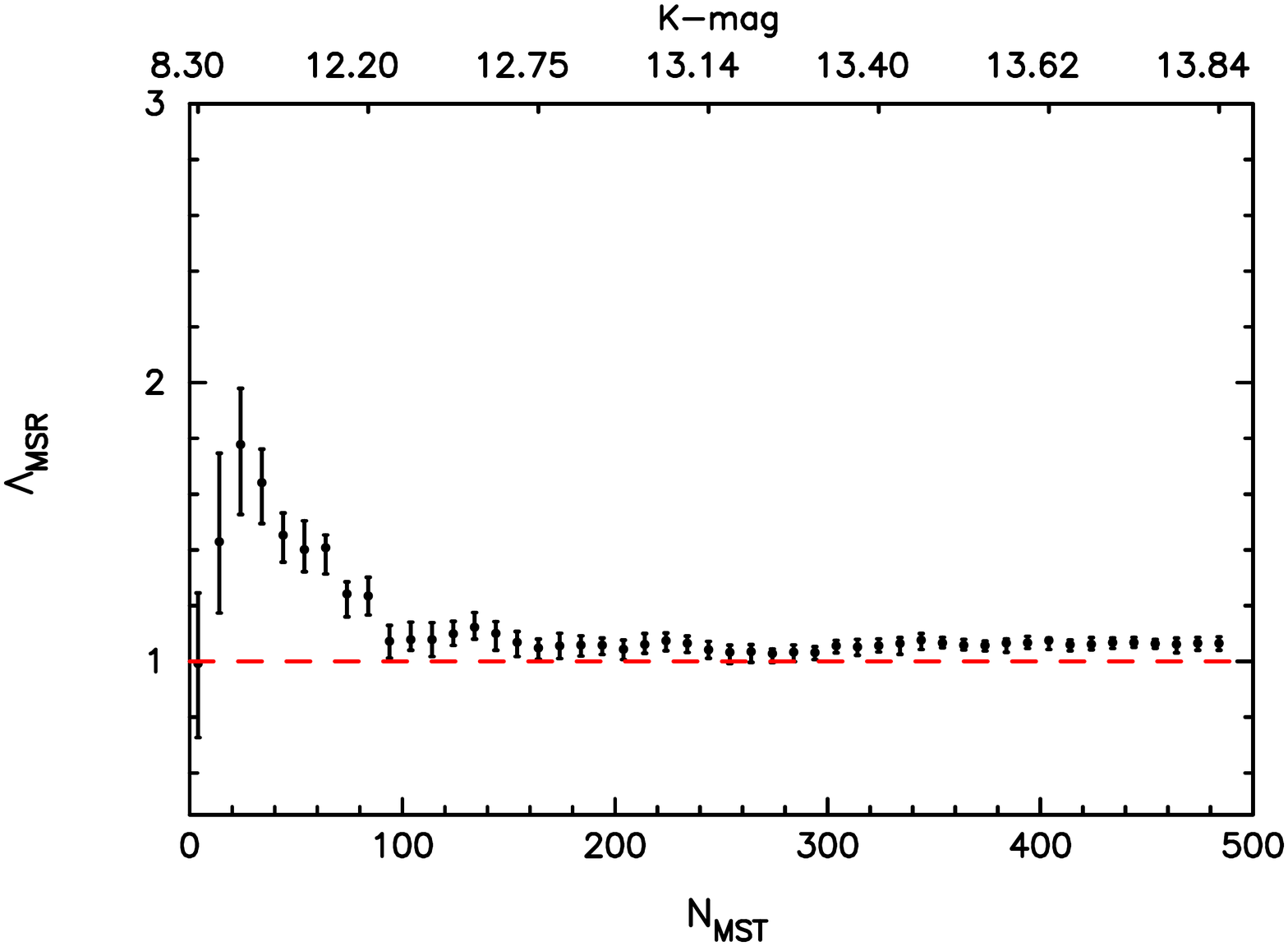}}}
\hspace*{0.2cm}\subfigure[]{\label{fig:Tr14_kuhn-c}\rotatebox{0}{\includegraphics[trim=1cm 0cm 5cm 1cm,scale=0.27]{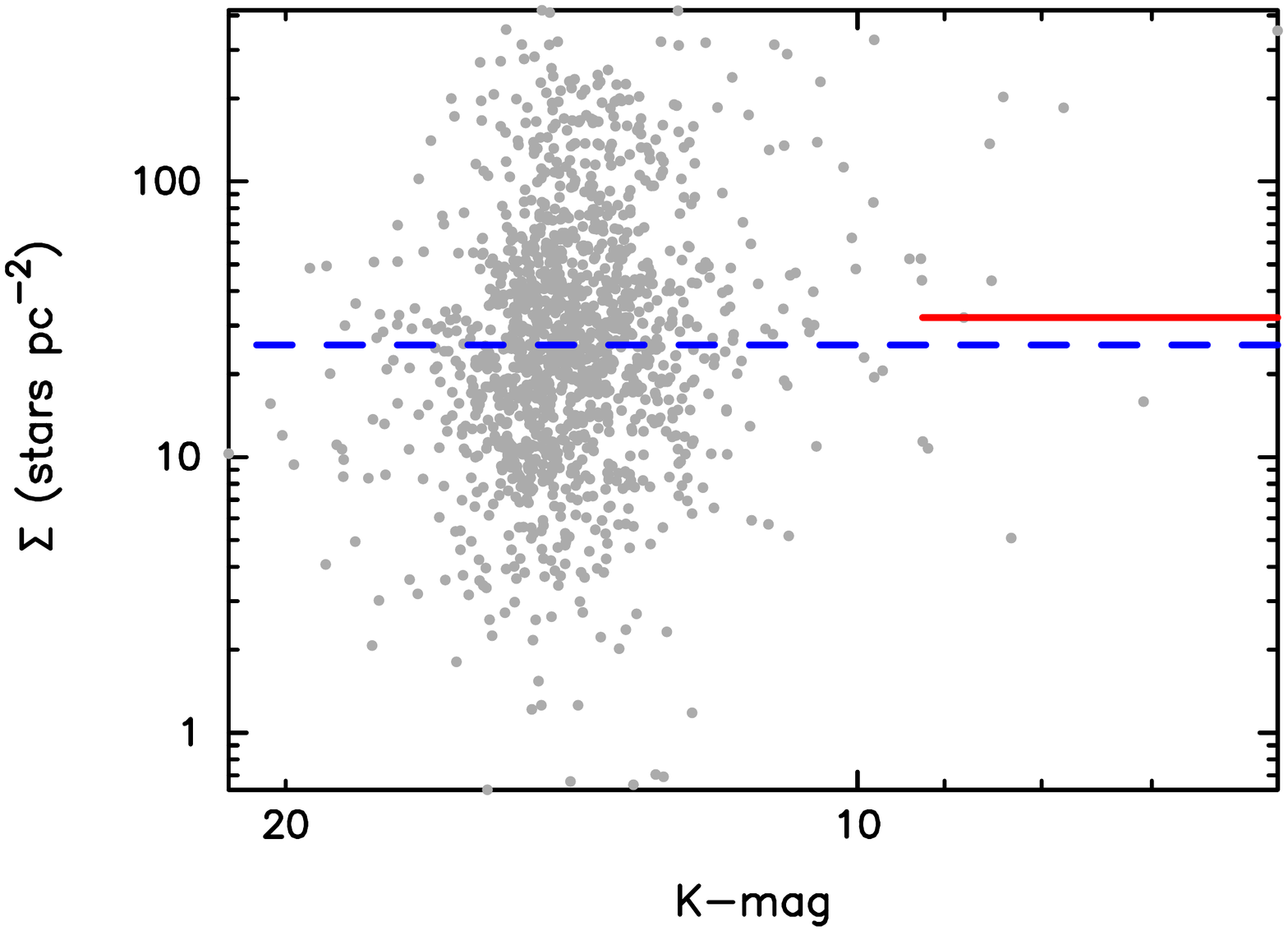}}}  
\caption[bf]{Structural analysis of Tr14 using the dataset defined in \citet{kuhn2014}. In panel (a) the brightest stars are shown by the red points. In panel (b) we show the $\Lambda_{\rm MSR}$ ratio as a function of the $N_{\rm MST}$ brightest stars.  The K-band magnitude of the least bright object enclosed in a sample of $N_{\rm MST}$ stars is indicated on the top axis. $\Lambda_{\rm MSR} = 1$ (no mass segregation) is shown by the dashed red line. In panel (c) we show the surface density $\Sigma$ for each star as a function of its K-band magnitude. The median surface density for the Carina region is shown by the blue dashed line, and the median surface density for the ten brightest (OB) stars is shown by the solid red line.}
\label{fig:Tr14_kuhn}
  \end{center}
\end{figure*}

In Fig.~\ref{fig:Tr14_kuhn} we show the positions of the brightest stars with respect to the other stars (panel a). Again, several of these are in binary systems. We repeat our calculation for the mass segregation ratio $\Lambda_{\rm MSR}$ as a function of the $N_{\rm MST}$ stars in the sample in panel (b). As in the compilation sample (Fig.~\ref{fig:Tr14_comp-c}) there is some deviation from $\Lambda_{\rm MSR} = 1$ for the brightest 60 stars (though not for the brightest 4 stars, see the leftmost datapoint in panel (b)).

Despite the very different samples, the overall shape of the distribution of $\Lambda_{\rm MSR}$ is very similar between the two samples. The slightly elevated  $\Lambda_{\rm MSR}$ ratio ($1 < \Lambda_{\rm MSR} < 2$) can be caused by stochastic populating of a random spatial distribution \citep{parker_goodwin_2015} and may not indicate a truly different spatial distribution for the most massive stars. 
It is possible that this type of phenomena is responsible for the spatial clustering of the massive stars determined by \citet{buckner2019}. 

\subsection{Tr16}

\begin{figure*}
  \begin{center}
\hspace*{-1.5cm}\subfigure[]{\label{fig:Tr16_comp-a}\rotatebox{0}{\includegraphics[trim=1cm 0cm 5cm 1cm,scale=0.4]{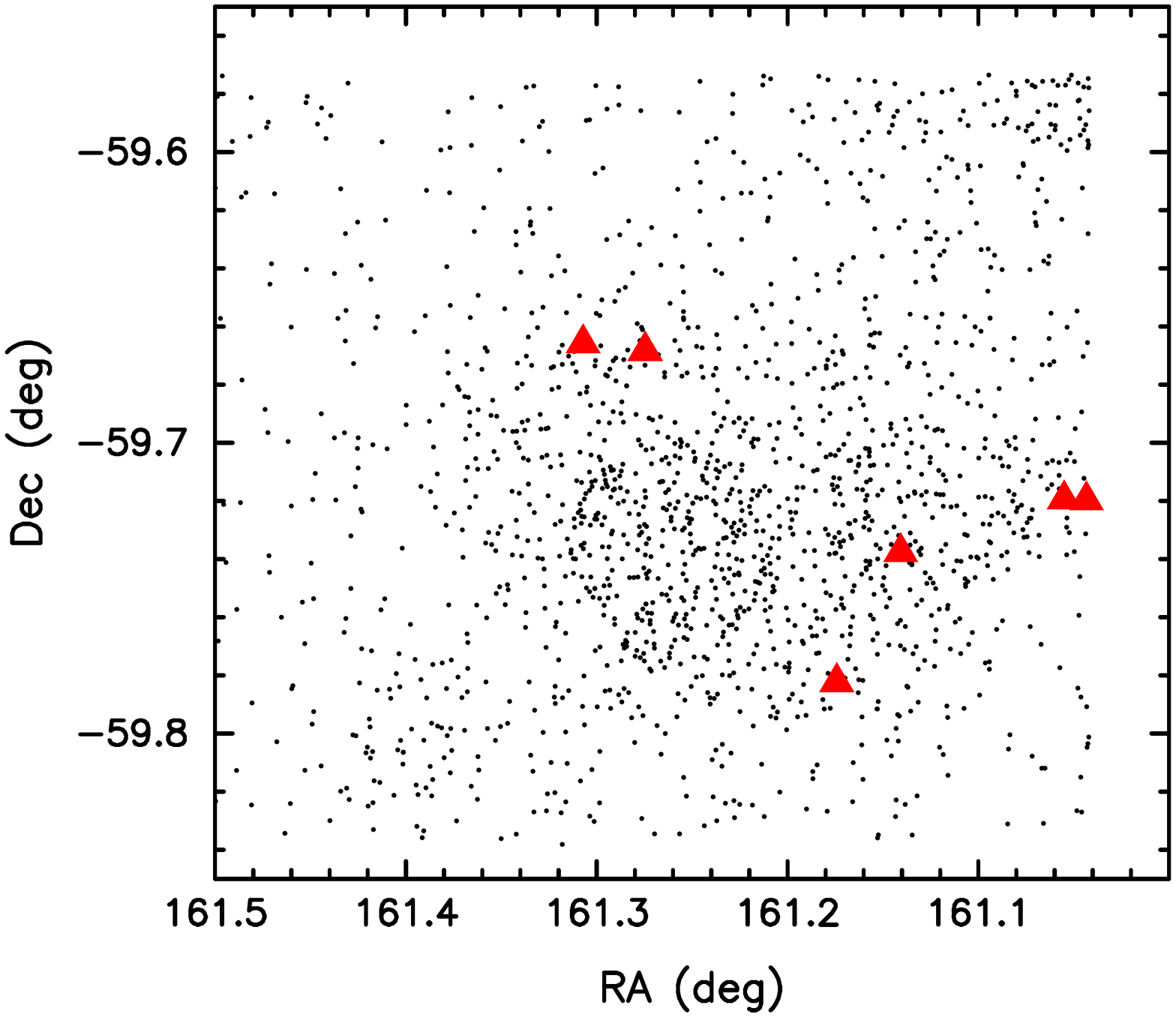}}}
\subfigure[]{\label{fig:Tr16_comp-b}\rotatebox{0}{\includegraphics[trim=1cm 0cm 5cm 1cm,scale=0.4]{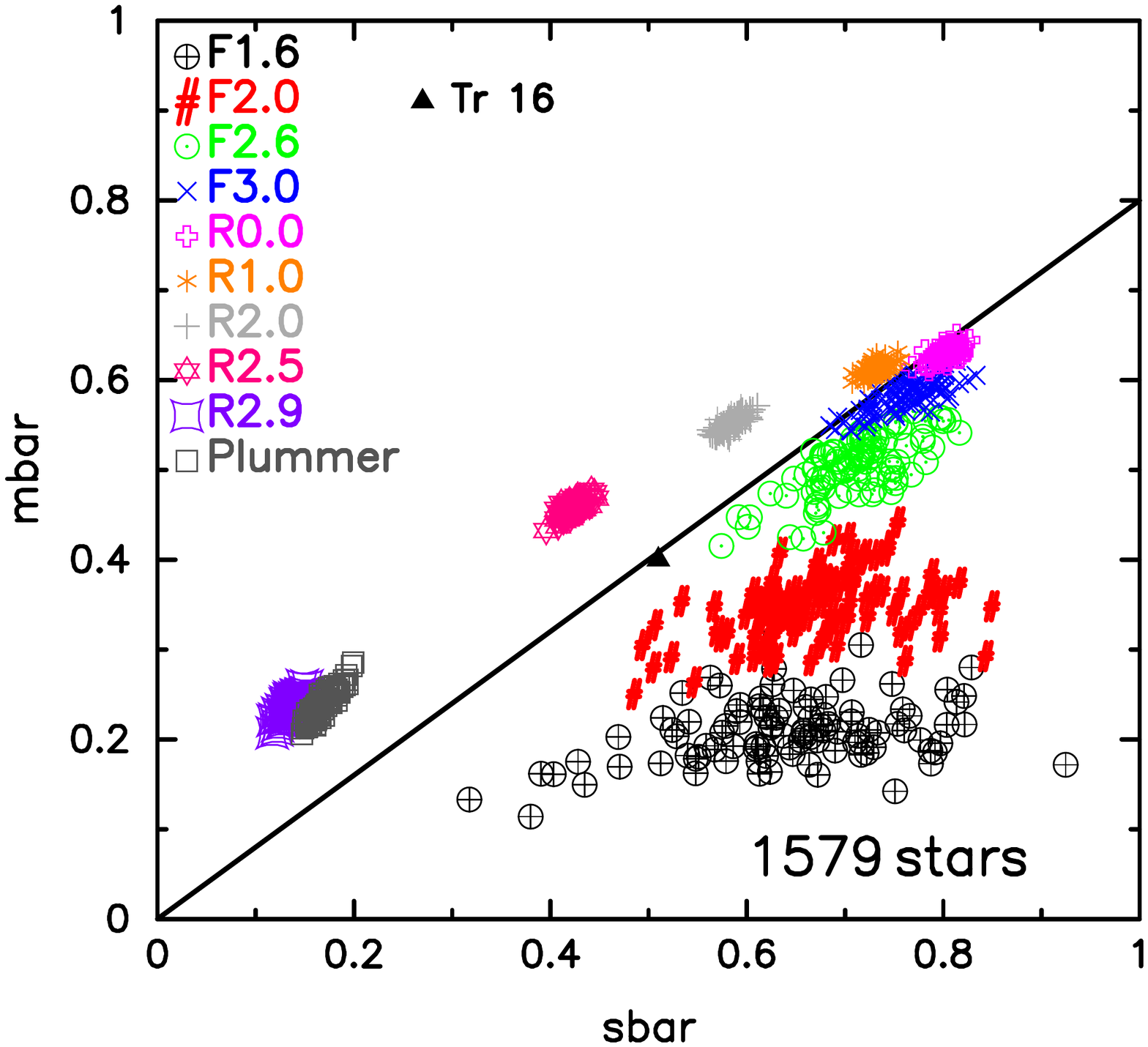}}} 
\hspace*{-0.85cm}\subfigure[]{\label{fig:Tr16_comp-c}\rotatebox{0}{\includegraphics[trim=2cm 0cm 5cm 1cm,scale=0.37]{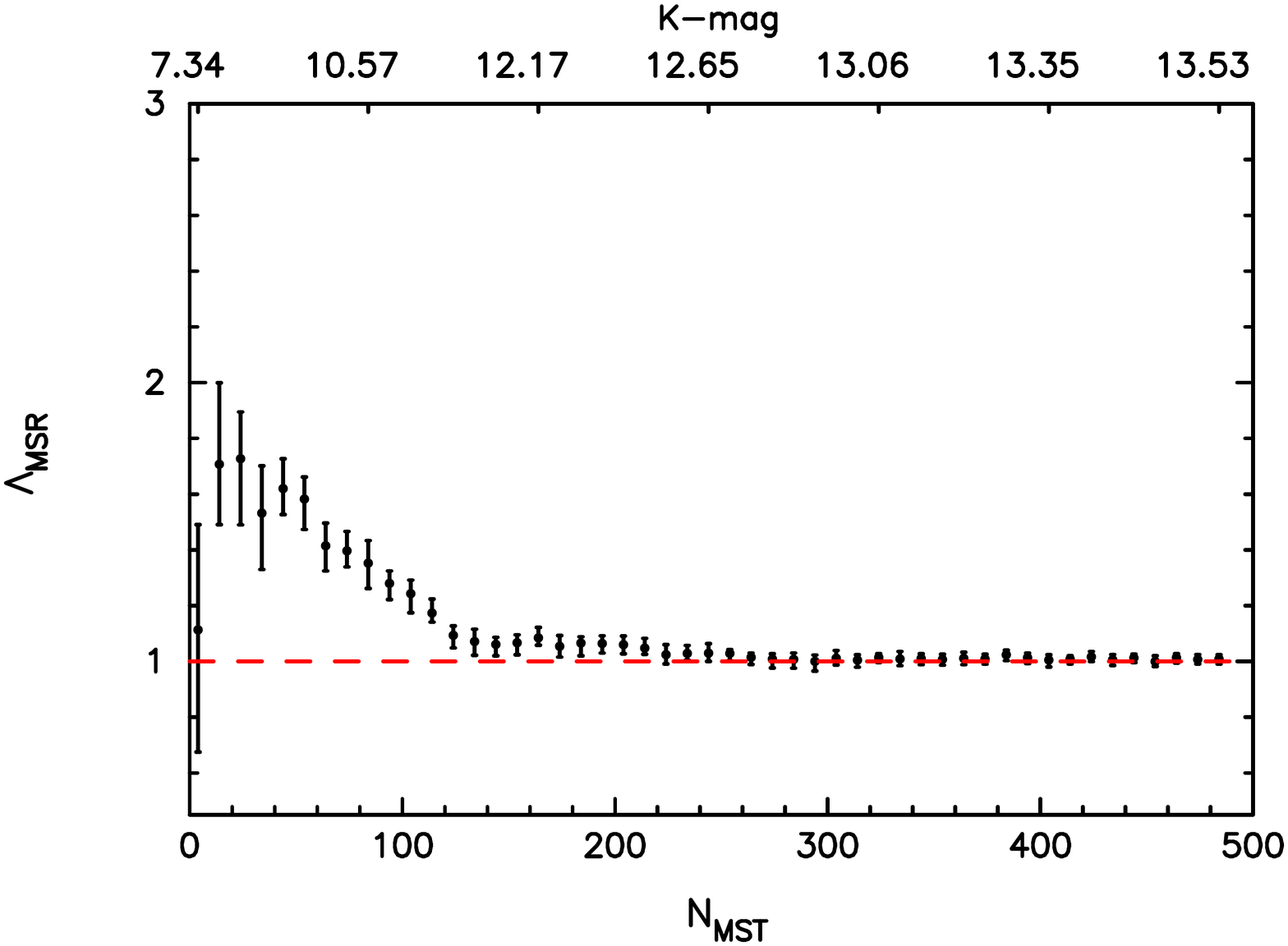}}}
\hspace*{0.5cm}\subfigure[]{\label{fig:Tr16_comp-d}\rotatebox{0}{\includegraphics[trim=1cm 0cm 5cm 1cm,scale=0.4]{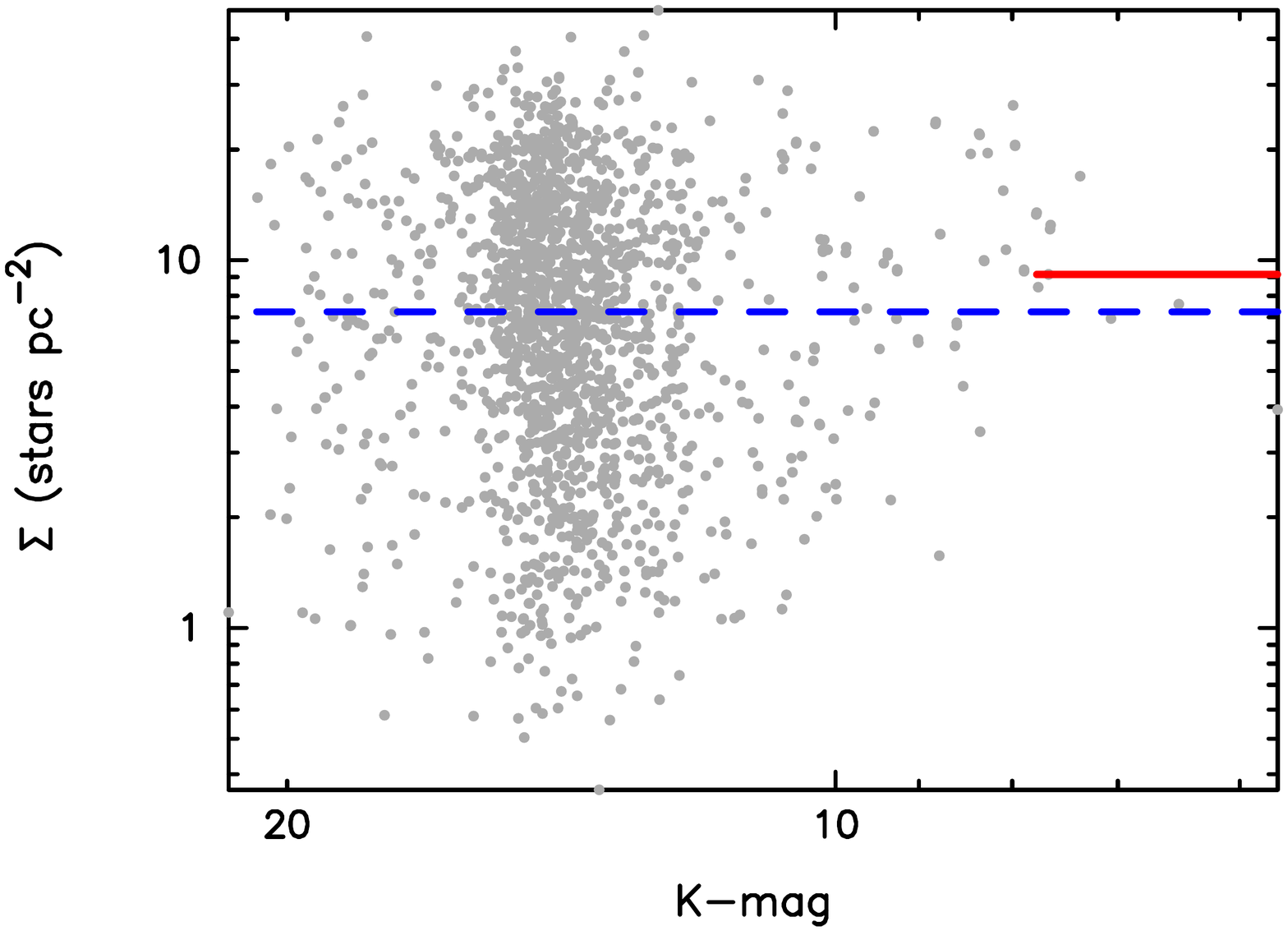}}}
\caption[bf]{Structural analysis of the Trumpler 16 dataset. In panel (a) the brightest stars are shown by the red points. In panel (b) we show the \citet{cartwright2009} plot, which plots the $\bar{m}$ and $\bar{s}$ components used to calculate the $\mathcal{Q}$-parameter against each other. The datum for Tr14 is shown by the solid black triangle, and for reference we show 100 realisations each of synthetic star-forming regions with various fractal dimensions, $D$,  (from a high degree of substructure [$D = 1.6$] to smooth [$D=3.0$] -- indicated as F1.6 - F3.0 in panel (b)), or centrally concentrated regions with different density profiles (uniform [$n \propto r^0$] to very centrally concentrated [$n \propto r^{-2.9}$] -- indicated as R0.0 - R2.9 in panel (b), as well as a Plummer profile). In panel (c) we show the $\Lambda_{\rm MSR}$ ratio as a function of the $N_{\rm MST}$ brightest stars.  The K-band magnitude of the least bright object enclosed in a sample of $N_{\rm MST}$ stars is indicated on the top axis.  $\Lambda_{\rm MSR} = 1$ (no mass segregation) is shown by the dashed red line. In panel (d) we show the surface density $\Sigma$ for each star as a function of its K-band magnitude. The median surface density for the Carina region is shown by the blue dashed line, and the median surface density for the ten brightest (OB) stars is shown by the solid red line.}
\label{fig:Tr16_comp}
  \end{center}
\end{figure*}

In Fig.~\ref{fig:Tr16_comp} we present our compilation data for Tr16. The positions of the ten brightest stars are shown by the red triangles in panel (a), some of which are in close binary systems. 

The \citet{cartwright2004} $\mathcal{Q}$-parameter is $\mathcal{Q} = 0.83$, where the mean branch length of the minimum spanning tree is $\bar{m} = 0.43$  and the mean length of the complete graph is $\bar{s} = 0.53$. These values are shown in the \citet{cartwright2009} plot in panel (b). Unlike for the full Carina region, or Tr14, this $\mathcal{Q}$-parameter is close to an idealised fractal geometry, with a small amount of spatial substructure. However, the $\mathcal{Q}$-parameter is only really powerful at distinguishing between a smooth or a substructured distribution; \citet{lomax2018} and Daffern-Powell \& Parker (subm.), find that it cannot be reliably used to trace the transition between these regimes, nor can it be used to infer the initial spatial distribution of a star-forming region. It can, however, be used as a proxy for the amount of dynamical evolution that has occurred in a star-forming region. 

We show the mass segregation ratio $\Lambda_{\rm MSR}$ as a function of the $N_{\rm MST}$ most massive stars in a chosen subset in Fig.~\ref{fig:Tr16_comp-c}. As with Tr14, the four most massive objects are not distributed differently to low-mass stars, but the next 50 stars show a marginally significant deviation from unity. As discussed above, that $\Lambda_{\rm MSR} < 2$ for all subsets means that this distribution may be consistent with a random distribution. 

Finally, we show the local surface density around each star against the K-band magnitude of the star in Fig.~\ref{fig:Tr16_comp-d}. The median surface density of the ten brightest stars is $\tilde{\Sigma}_{\rm 10} = 9$\,stars\,pc$^{-2}$, whereas the median surface density for the full sample is $\tilde{\Sigma}_{\rm all} = 7$\,stars\,pc$^{-2}$. We therefore conclude that the brightest stars in Tr16 are not in areas of higher than average surface density. 

\section{Limited dynamical evolution in Tr14 and Tr16}\label{s:discussion}

The evolution of cluster density with time has a huge affect on the fate of protoplanetary discs by regulating the intensity of the incident photoevaporative flux \citep{nicholson2019}. 
Direct comparison with simulations is difficult if the cluster density has evolved significantly since formation, as the dynamical history is difficult to reconstruct. 
Fortunately, structural analyses provide a way to constrain the dynamical history of the cluster.  

The metric used to define mass segregation in this paper ($\Lambda_{\rm MSR}$, see Section~\ref{ss:Lmsr}) suggests no mass segregation in either Tr14 or Tr16. 
Using a different statistical analysis, 
\citet{buckner2019} find evidence for mass segregation in Tr14 and Tr15, an older cluster located near the northeast edge of the larger Carina star-forming region. 
Different methods for measuring mass segregation often produce different and sometimes contradictory results \citep{parker_goodwin_2015}.
For clusters with smooth radial profiles and well-defined mass segregation like Tr15 \citep{wang2011}, most metrics agree. 
Regions with more (or ambiguous) substructure have more variation in local density measures, producing more statistical fluctuations in mass segregation estimators. 
In addition, the relatively small number of high-mass stars, even in rich clusters like Tr14 and Tr16, tends to exacerbate this sensitivity to statistical fluctuations.
Nevertheless, contradictory conclusions on mass segregation in Tr14 in the literature \citep{ascenso2007,sana2010,buckner2019} are consistent with our finding of weak or absent mass segregation, especially given that a low-level signature of mass segregation can be produced by random chance. 
In addition, many of the brightest objects in our samples are in binary systems with other OB stars. Therefore, the OB \emph{systems} are not mass segregated, but a modest mass segregation signal may occur due to binarity in the sample \citep{maschberger2011}.

We argue that the absence of mass segregation is evidence that both Tr14 and Tr16 are dynamically young. 
More dynamical processing would lead to more mass segregation and a smoother cluster profile. 
However, the $Q$-parameter is also inconclusive, suggesting that stars have not settled into a smooth radial distribution. 
Together, these metrics suggest a region that is dynamically young, with a density that is not significantly different from its primordial density \citep{parker2012}.

Little evidence for mass segregation, coupled with an inconclusive $Q$-parameter suggests that the cluster densities in Tr14 and Tr16 have been relatively consistent over time.
Assuming that the current density reflects the primordial density allows us to compare to simulations of cluster evolution -- and protoplanetary disc destruction -- for clusters of a given density. 
In the next section, we compare the predicted fraction of surviving protoplanetary discs to the observed percentage of disk-bearing near-IR excess sources in each cluster.

\section{Protoplanetary disc survival}\label{ss:disk_distro}

As exoplanet surveys increase the number of known terrestrial and potentially habitable planets, there is growing urgency to develop a comprehensive understanding of planet formation.
Environment is a key dimension, as observations and numerical simulations suggest that the local ecology significantly impacts the survival and enrichment of planet-forming discs. 
Many authors have considered the role of the cluster environment in disc survival and planet formation \citep[e.g.,][]{adams2004,adams2006,clarke2007,winter2018}.
Despite this, most detailed work on the impact of feedback on planet-forming discs comes from the ONC \citep[e.g.,][]{odell1994,johnstone1998,throop2005,mann2010,mann2014,eisner2018}. 

In the following discussion, we assume that all stars will form a disk as part of their evolution given the overwhelming evidence for flattened, rotating structures seen around stars of all masses during their early evolution \citep[e.g.,][]{williams2011,beltran2016,johnston2015,maud2018}. Metallicity may play an important role in the chemistry and sedimentation in the disc and the fraction of metals in the disc may evolve as gas is photoevaporated, increasing the dust-to-gas ratio \citep[e.g.,][]{throop2005}. While interesting, these effects are not the focus of our study. Instead, we use an empirical measurement (the near-IR excess) to identify sources that still have (hot) circumstellar dust. We compare this to models that take photoevaporation rates, surface density profiles, and disc masses from recent results in the literature. At lower metallicity, the impact of ionizing feedback may be enhanced as the gas is less able to self-shield, accelerating disc destruction. A thorough analysis of this possibility is beyond the scope of the paper, and we assume that the main factor in determining disc evolution is the cluster density and radiation field.

\citet{preibisch2011} report near-IR colors and magnitudes for thousands of X-ray-emitting YSOs in Carina, probing to the X-ray detection limit of $\sim 0.5-1$~M$_{\odot}$. 
Relatively few sources meet their criteria for near-IR excess emission that indicates the presence of a circumstellar disk.
Derived disc fractions are similar in Tr14 ($9.7 \pm 0.8$)\% and Tr16 ($6.9 \pm 1.2$)\%. 
In both cases, these fractions are lower than those measured in small nearby clusters of similar age \citep[see][]{preibisch2011_hawki}.

A few studies have found a spatial trend in the distribution of IR-excess sources with a higher fraction detected at larger distances from the high-mass stars (e.g., \citealt{balog2007}, \citealt{guarcello2007}, although see \citealt{roccatagliata2011} and \citealt{busquet2019}). 
Disc masses, estimated from millimeter continuum emission, also appear to be higher in sources located further from the high-mass stars 
\citep[e.g.,][]{mann2010,mann2014,ansdell2017,eisner2018}. 
However, \citet{richert2015} argue that there is no evidence for a spatial stratification in a sample of 6 high-mass star forming regions.
Instead, they argue that both disk-bearing and disk-less sources appear under abundant, suggesting an observational bias.

\begin{figure}
  \centering
  $\begin{array}{c}
  \includegraphics[trim=0mm 0mm 0mm 0mm,angle=0,scale=0.575]{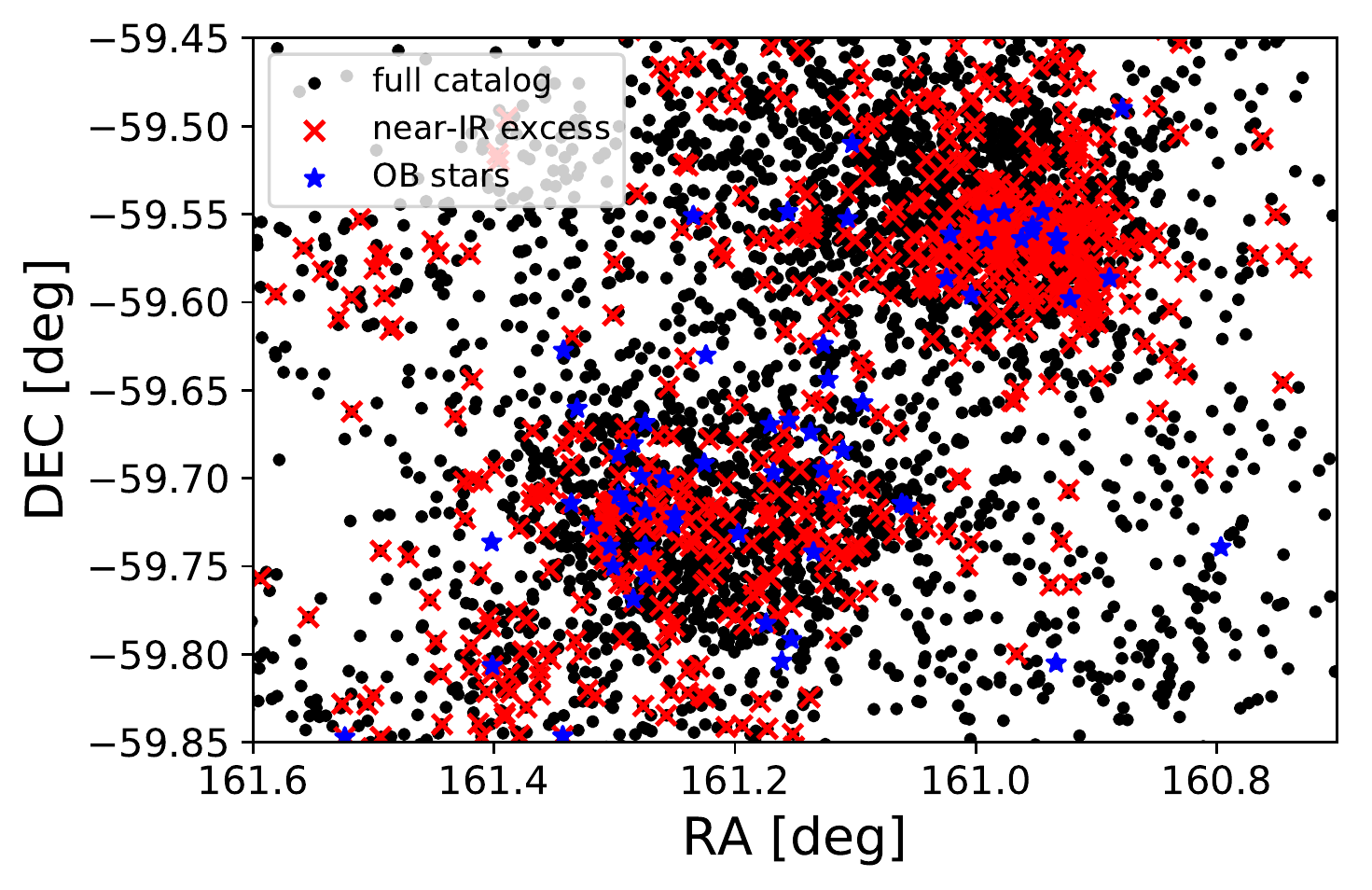} \\
  \includegraphics[trim=0mm 0mm 0mm 0mm,angle=0,scale=0.575]{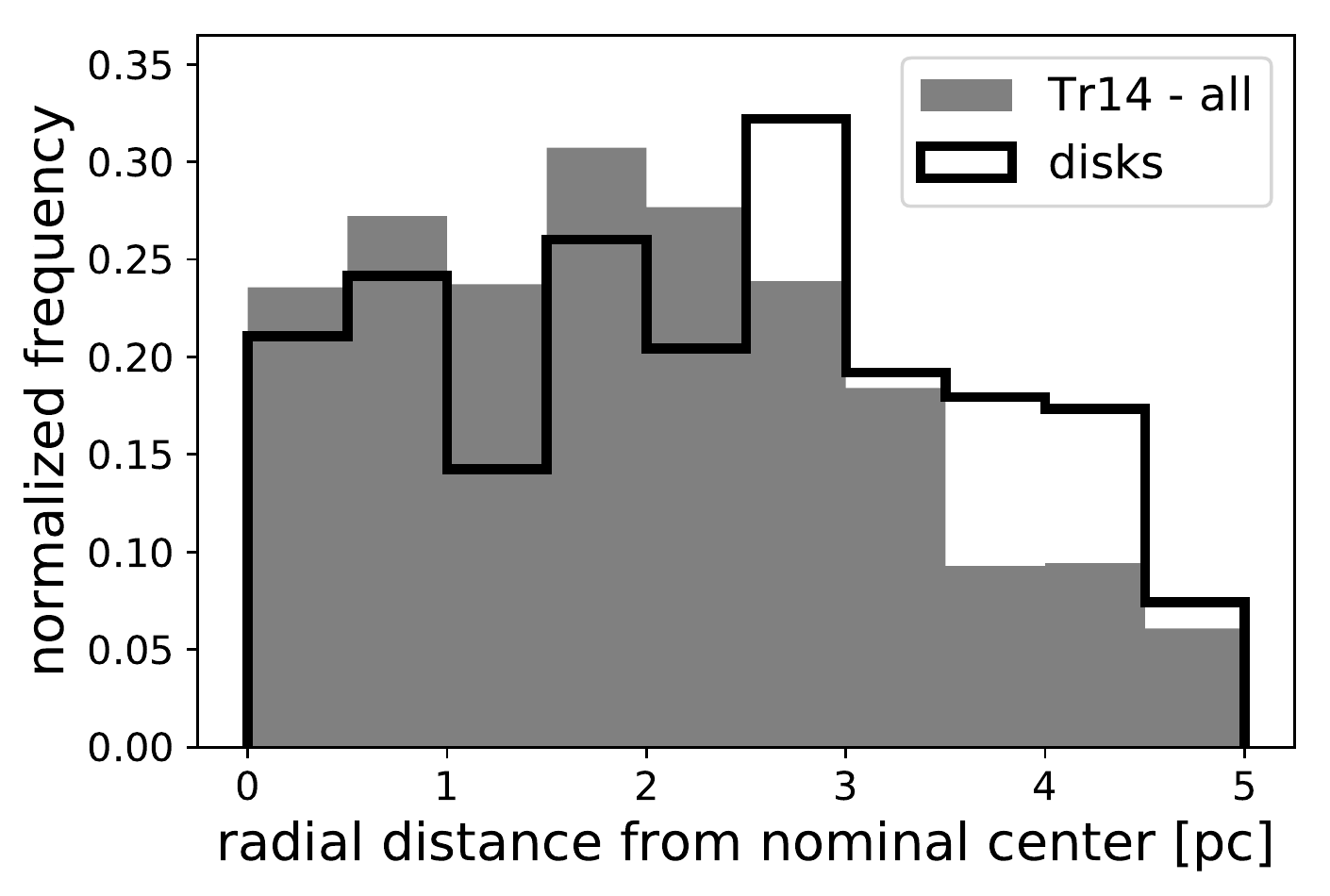} \\
  \includegraphics[trim=0mm 0mm 0mm 0mm,angle=0,scale=0.575]{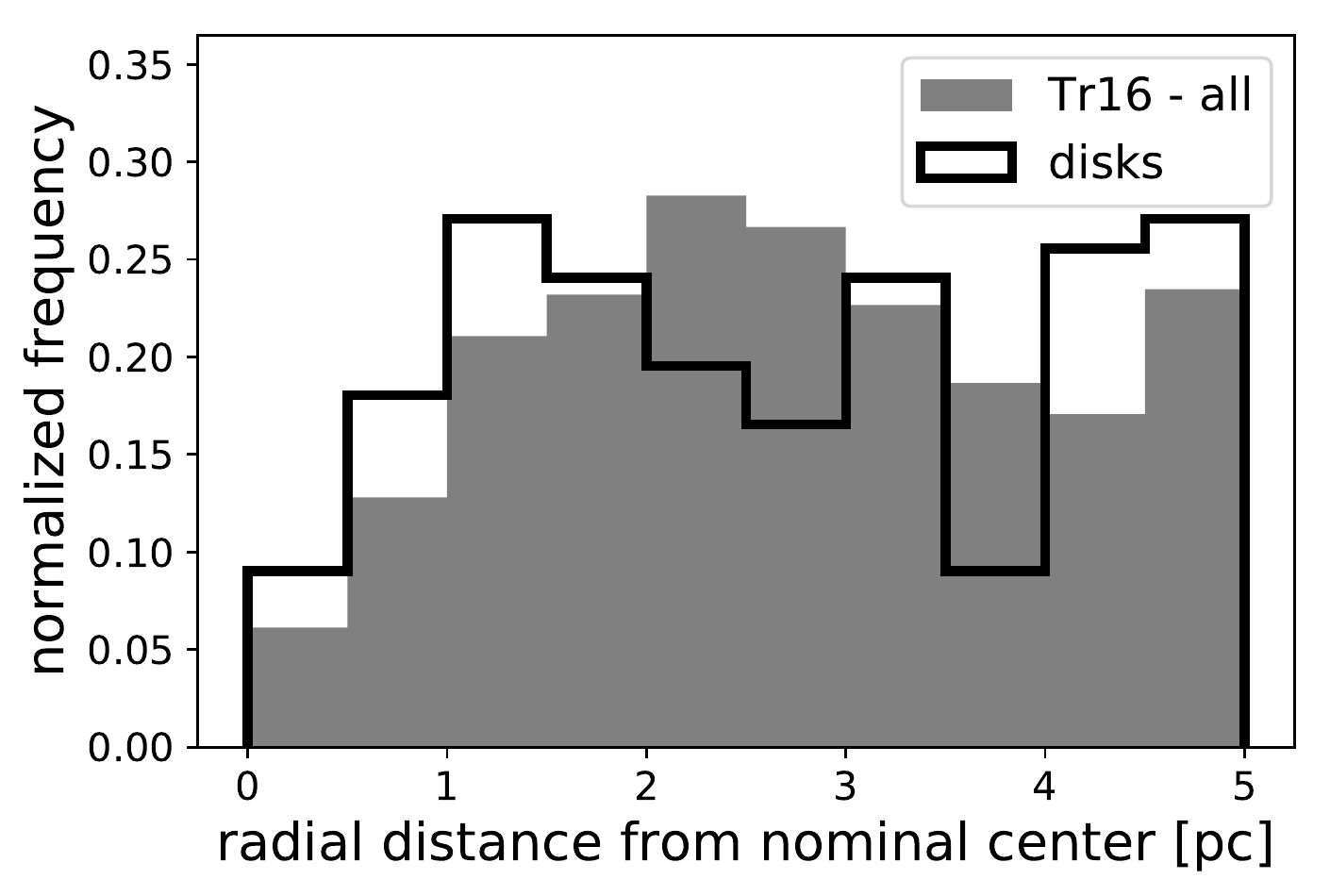} \\
  \end{array}$
\caption{
\textit{Top:} The distribution of near-IR-excess (disk-bearing) sources \citep[red crosses; data from][]{preibisch2011} is shown compared to all point sources used in this analysis (black dots). High-mass stars are denoted with blue stars. 
\textit{Middle:} Histograms comparing the radial distribution of stars in Tr14 (gray) with the distribution of IR-excess sources (black outline). 
\textit{Bottom:} Histograms comparing the radial distribution of stars and IR-excess sources in Tr16. 
}\label{fig:discs} 
\end{figure}

We show the distribution of low-mass stars with a near-IR excess from \citet{preibisch2011} in Figure~\ref{fig:discs}.  
There is no obvious spatial structure in the distribution of IR-excess sources, in agreement with the findings of \citet{richert2015} for Carina. 
To quantify this, we compare the projected radial distribution of stars with and without an IR-excess in each cluster. 
A K-S test comparing the two populations returns the same probability (p=0.68) for both clusters, suggesting no spatial dependence in the disc distribution.

In fact, it is not clear that we should expect a structured distribution of disc-bearing sources.
Both clusters have multiple high-mass stars distributed throughout the clusters (neither cluster is mass-segregated, see Section~\ref{s:discussion}). 
More importantly, clusters are not static and the separation between high- and low-mass stars is not constant. 
A star whose disc has been evaporated will not stay in the same place for long, and will dynamically mix with stars that retain some or all of their discs. Therefore, we do not expect a strong correlation between disc mass and distance from any ionising sources (i.e.\,\,massive stars). 

Furthermore, observations that do show a dependence of disc mass on distance from the massive stars are by definition two-dimensional projections of a three-dimensional distribution. Parker et al.\,\,(in prep.) show that projection effects will significantly hamper any interpretation of the spatial distributions of stars with and without discs.

Most studies on the destruction of protoplanetary discs agree that FUV photoevaporation is the dominant factor in this destruction process \citep{storzer1999,scally_clarke2001,adams2004,haworth2018,winter2018,nicholson2019}. FUV leads to photoevaporative mass-loss rates on the order of 0.2\,M$_\odot$\,Myr$^{-1}$ \citep{scally_clarke2001,nicholson2019}, implying that a 1\,M$_\odot$ star with a disc that is initially 10\,per cent of its mass could expect to lose the gas content from this disc on timescales of 0.5\,Myr. 

The crossing time in a star-forming region is the time taken for a star to traverse the spatial extent of the region, and is used as a proxy to estimate how much dynamical evolution has taken place. If the age of the star-forming region exceeds the crossing time by a factor of several, then the region is likely to be dynamically old, and the stars will be dynamically mixed. Dense star-forming regions ($> 10^3$~stars\,pc$^{-3}$) have crossing times of order 0.1\,Myr, whereas low-density regions ($<10$\,stars\,pc$^{-3}$) will have crossing times of order several Myr.

Given the lack of strong correlation between near-IR excess and the spatial distribution of stars in Carina, 
it is tempting to conclude that the region is dynamically old and the discs have been destroyed by the ambient radiation field. 
However, the statistical diagnostics we apply to Tr14 and Tr16 do not indicate mass segregation or highly substructured clusters.
We take this as evidence that the cluster density has not evolved significantly since birth, allowing us to compare to numerical simulations that predict disc survival rates. 
\citet{nicholson2019} predict that a cluster with a density $\sim 10$\,stars~pc$^{-2}$ will have a remaining disc fraction of $\sim 10$\% after $\sim 3$~Myr.  
For higher densities, $\sim 100$\,stars~pc$^{-2}$, discs will be dissipated faster, with $\sim 10$\% remaining after $\sim 1-2$~Myr.
We note that these disc fractions are remarkably similar to what \citet{preibisch2011} report for Tr16 and Tr14, respectively, assuming that both clusters represent a single-age population. 
In reality, most star-forming regions show some evidence for age spreads, although we note that stars that are significantly older or younger tend to be in spatially distinct portions of the larger star-forming complex \citep[see, e.g.,][]{smith2008,getman2014}.

Only surviving protoplanetary discs will be enriched with the radioactive isotopes synthesized and ejected during the deaths of the highest mass cluster members. 
\citet{lichtenberg2016} estimated the enrichment distribution and radiogenic heating from  $^{26}$Al in high-mass stellar clusters and found a broad distribution of expected values, including those consistent with the calculated heat budget for the interior of Earth. 
Their simulated star-forming regions were at least an order of magnitude more dense (1000\,stars\,pc$^{-3}$) than the present-day value for either cluster in Carina, preventing a direct comparison with their derived enrichment levels. \citet{nicholson2017} performed simulations of lower-density star-forming regions of comparable density to Carina in order to determine the number of stars that could be enriched; however, \citet{nicholson2017} did not perform the full internal heating calculations and it remains an open question as to whether these low-density clusters could produce the observed levels of $^{26}$Al.

While we cannot compare directly with simulations of disc enrichment, we note two potential benefits for low-mass stars born in high-mass clusters like Tr14 and Tr16. 
Compared to smaller regions, both Tr14 and Tr16 have higher mass stars that will explode as supernovae earlier, possibly before the destruction of the remaining discs. 
More importantly, more massive stars (M$>25$~M$_{\odot}$) synthesize and eject $^{26}$Al during their pre-supernova evolution \citep{limongi2006}, enriching the local interstellar medium earlier ($\sim 3$~Myr) than supernovae \citep{voss2009}.

Observations of the 1.8~MeV decay line of $^{26}$Al show that it correlates with OB associations \citep{knoedlseder1996}. 
The derived abundance in the Carina region exceeds that which can plausibly be produced by supernovae alone, strongly suggesting additional enrichment from winds \citep{voss2012}. 
The estimated mass of $^{26}$Al currently in Carina is
0.004--0.009~M$_{\odot}$, corresponding to a mass fraction of $\sim 10^{-9} - 10^{-8}$ \citep[using the mass of gas and dust from][]{preibisch2012}. 
The lower bound of this estimate overlaps with the high end of the Galactic average estimate from \citet{lugaro2018}. 
At the high end, the abundance in Carina is an order of magnitude higher than the Galactic average, and thus much closer to the value inferred for the early Solar System \citep[see][]{jacobsen2008,lugaro2018}.

\section{Conclusions}\label{s:conclusions}
Stars and planets often form in the context of a larger clustered environment where feedback from nearby high-mass stars will affect the survival and enrichment of planet-forming discs, particularly around low-mass stars. 
The dynamical evolution of the cluster plays a critical role by regulating the amount of time that low-mass stars spend subject to disc-destroying ionizing radiation. 
We present a structural analysis to constrain the dynamical histories of Tr14 and Tr16, two high-mass clusters in the Carina Nebula. 
Neither cluster shows evidence for mass segregation and the $\mathcal{Q}$-parameter, a diagnostic for substructure, is inconclusive. 
We take this as evidence for limited dynamical evolution in both clusters. 
This allows us to compare to the disc fractions predicted to survive in clusters of these densities by \citet{nicholson2019}. 
The predicted surviving disc fractions are $\sim 10$\%, remarkably similar to those reported in Tr14 and Tr16 by \citet{preibisch2011}, providing further evidence of the important role of the cluster environment in shaping planet-forming discs.


\section*{Acknowledgements}
This project has received funding from the European Union's Horizon 2020 research and innovation programme under the Marie Sk\'{l}odoska-Curie grant agreement No.\ 665593 awarded to the Science and Technology Facilities Council. RJP acknowledges support from the Royal Society in the form of a Dorothy Hodgkin Fellowship. This research has made use of the SIMBAD database, operated at CDS, Strasbourg, France.

\bibliographystyle{mnras}
\bibliography{bibliography_mrr}





\bsp	
\label{lastpage}
\end{document}